\documentclass[manuscript,screen]{acmart}

\AtBeginDocument{%
  \providecommand\BibTeX{{%
    \normalfont B\kern-0.5em{\scshape i\kern-0.25em b}\kern-0.8em\TeX}}}

\acmJournal{JACM}
\acmVolume{37}
\acmNumber{4}
\acmArticle{111}
\acmMonth{8}

\acmPrice{15.00}
\acmISBN{978-1-4503-XXXX-X/18/06}

\usepackage{multirow}
\usepackage{xcolor}
\usepackage{colortbl}
\usepackage{pifont}
\usepackage[tikz]{bclogo}
\usepackage[framemethod=tikz]{mdframed}
\usepackage{lipsum}
\usepackage{subfloat}
\usepackage[subrefformat=parens,labelformat=parens]{subfig}
\usepackage{enumitem}
\usepackage{xspace}
\usepackage{makecell}
\usepackage{array}
\usepackage{longtable}

\newcolumntype{L}[1]{>{\raggedright\let\newline\\\arraybackslash\hspace{0pt}}m{#1}}
\newcolumntype{C}[1]{>{\centering\let\newline\\\arraybackslash\hspace{0pt}}m{#1}}
\newcolumntype{R}[1]{>{\raggedleft\let\newline\\\arraybackslash\hspace{0pt}}m{#1}}

\begin{document}
\newcommand{\xzl}[1]{\textcolor{purple}{(Xuanzhe: #1)}}
\newcommand{\gdd}[1]{\textcolor{red}{#1} }
\newcommand{\czp}[1]{\textcolor{blue}{czp: #1} }
\newcommand{\xin}[1]{\textcolor{blue}{Xin: #1} }
\newcounter{finding}
\newcommand{\revise}[1]{\textcolor{black}{#1}}

\newcommand{\finding}[1]{\refstepcounter{finding}
  \vspace{2.3mm}
 \begin{mdframed}[linecolor=gray,roundcorner=12pt,backgroundcolor=gray!15,linewidth=3pt,innerleftmargin=2pt, leftmargin=0cm,rightmargin=0cm,topline=false,bottomline=false,rightline = false]
  \textbf{Finding \arabic{finding}:} #1
 \end{mdframed}
 \vspace{2.3mm}
}
\newcommand*\blackcircled[1]{\tikz[baseline=(char.base)]{
		\node[shape=circle,fill=black,draw,text=white,inner sep=1pt] (char) {#1};}}
\captionsetup[figure]{font=bf,skip=10pt}
\captionsetup[table]{font=bf,skip=10pt}
\newcommand{\distance}{3pt}
\setlength{\textfloatsep}{1pt}
\setlength{\floatsep}{\distance}
\setlength{\intextsep}{\distance}
\setlength{\dbltextfloatsep}{\distance} 
\setlength{\dblfloatsep}{\distance} 
\newcommand{\sysname}{\texttt{DTClinic}\xspace}

\title{Rise of Distributed Deep Learning Training in the Big Model Era: From a Software Engineering Perspective}

\author{Xuanzhe Liu}
\affiliation{
    \institution{Peking University}
    \city{Beijing}
    \country{China}}
\email{liuxuanzhe@pku.edu.cn}

\author{Diandian Gu}
\affiliation{
    \institution{Peking University}
    \country{China}}
\email{gudiandian1998@pku.edu.cn}

\author{Zhenpeng Chen}
\affiliation{
    \institution{University College London}
    \city{London}
    \country{UK}}
\email{zp.chen@ucl.ac.uk}

\author{Jinfeng Wen}
\affiliation{
    \institution{Peking University}
    \city{Beijing}
    \country{China}}
\email{jinfeng.wen@stu.pku.edu.cn}

\author{Zili Zhang}
\affiliation{
    \institution{Peking University}
    \city{Beijing}
    \country{China}}
\email{zzlcs@pku.edu.cn}

\author{Yun Ma}
\affiliation{
    \institution{Peking University}
    \city{Beijing}
    \country{China}}
\email{mayun@pku.edu.cn}

\author{Haoyu Wang}
\affiliation{
    \institution{Huazhong University of Science and Technology}
    \country{China}}
\email{haoyuwang@hust.edu.cn}

\author{Xin Jin}
\affiliation{
    \institution{Peking University}
    \city{Beijing}
    \country{China}}
\email{xinjinpku@pku.edu.cn}

\renewcommand{\shortauthors}{Liu et al.}

\begin{abstract}
\revise{Deep learning (DL) has become a key component of modern software. In the ``\textit{big model}'' era, the rich features of DL-based software (i.e., DL software) substantially rely on powerful DL models, e.g., BERT, GPT-3, and the recently emerging GPT-4, which are trained on the powerful cloud with large datasets. Hence, training effective DL models has become a vital stage in the whole software lifecycle. When training deep learning models, especially those big models, developers need to parallelize and distribute the computation and memory resources amongst multiple devices (e.g., a cluster of GPUs) in the training process, which is known as \textit{distributed deep learning training}, or \textit{\textbf{distributed training}} for short. However, the unique challenges that developers encounter in distributed training process have not been studied in the software engineering community. Given the increasingly heavy dependence of current DL-based software on distributed training, this paper aims to fill in the knowledge gap and presents the first comprehensive study on developers' issues in distributed training. 
To this end, we focus on popular DL frameworks that support distributed training (including TensorFlow, PyTorch, Keras, and Horovod) and  analyze 1,131 real-world developers' issues about using these frameworks reported on Stack Overflow and GitHub.
We construct a fine-grained taxonomy consisting of 30 categories regarding the fault symptoms and summarize common fix patterns for different symptoms. We find that : (1) many distributed-specific faults and non-distributed-specific faults inherently share the same fault symptoms, making it challenging to debug; (2) most of the fault symptoms have frequent fix patterns; (3) about half of the faults are related to system-level configurations. Based on the results, we suggest actionable implications on research avenues that can potentially facilitate the  distributed training to develop DL-based software, such as focusing on the frequent and common fix patterns when designing testing or debugging tools, developing efficient testing and debugging techniques for communication configuration along with the synthesis of network configuration analysis, designing new multi-device checkpoint-and-replay techniques to help reproduction, and designing serverless APIs for cloud platforms.}
\end{abstract}


\maketitle
\section{Introduction}
\label{sec:intro}
\begin{figure}[t]
    \centering
    \includegraphics[width=0.55\linewidth]{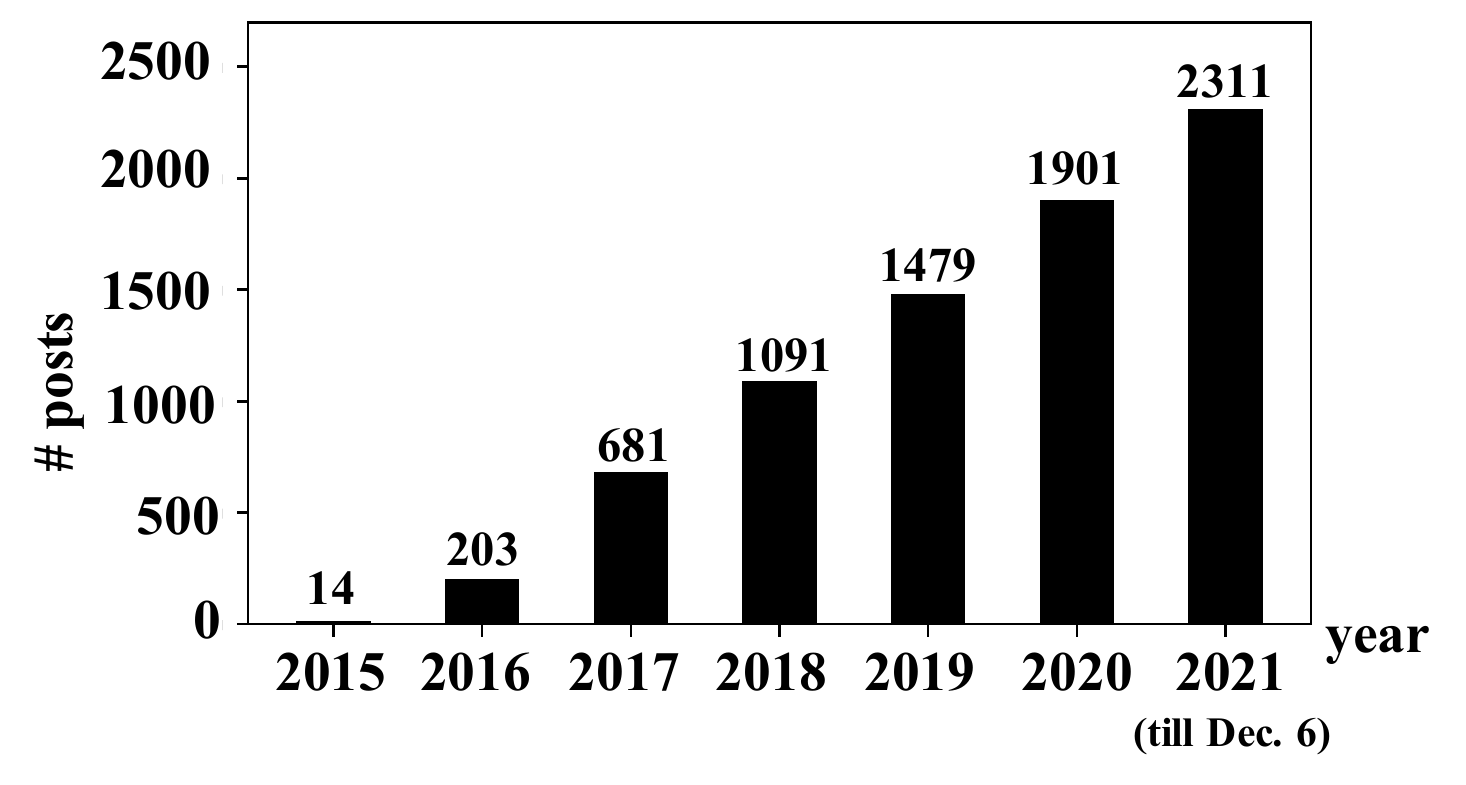}
        \vspace*{-2ex}
    \caption{The number of distributed-training-related posts on SO \revise{until each year}.}
    \label{fig:post}
\end{figure}

Deep learning (DL) has been a key component in modern software, ranging from supporting daily activities (e.g., speech-to-text~\cite{BerardPSB16}) to safety-critical tasks (e.g., autonomous vehicles~\cite{ChenSKX15}). The rich features of these DL-based software applications (i.e., DL software) increasingly rely on powerful ``big'' DL models. 
To increase the accuracy of DL models, on the one hand, a substantial volume of training data is required; 
on the other hand, the DL model architectures become more and more complex, e.g., BERT large~\cite{DevlinCLT19} with 340 million parameters and GPT-3~\cite{gpt3} with 175 billion parameters. 
As data increases in volume and DL models in complexity, the computational intensity and memory demand of DL increase proportionally~\cite{Ben-NunH19}. It is reported that the computation demand of DL training grows at a speed of 35$\times$ every 18 months~\cite{openai}.
As a result, developers have no other options but to parallelize and distribute computation and memory to multiple devices (e.g., GPUs and servers) during the training process of DL models, i.e., \emph{distributed training}. Distributed training has drawn a lot of attention from the research community~\cite{HeZRS16, DevlinCLT19, JeonVPQXY19, HazelwoodBBCDDF18, MLaaS, JiangZLYCG20, PengZCBYLWG19, narayanan2019pipedream}, and gained significant considerations in popular AI frameworks like TensorFlow~\cite{tensorflow}, PyTorch~\cite{pytorch}, Horovod~\cite{horovod}, etc. It is also observed that distributed training draws a lot of attention from software developers. \revise{Fig.~\ref{fig:post} presents the cumulative number of the distributed-training-related posts on Stack Overflow (SO) over the years to 2021. We can see the overall trend of continuous interest from developers in this topic.}


Compared to non-distributed training, distributed training has its unique features and challenges. Distributed training is performed in more complex environment settings, requiring protocols (e.g., for communication among devices) and algorithms (e.g., for collaborations among devices).
As a result, developers inevitably encounter a variety of issues about distributed training in practice and these issues are frequently asked on developers’ Q\&A forums. 
For example, some developers find it difficult to configure communication between multiple devices that participate in distributed training~\cite{commconfig} and complain that they cannot achieve the expected training speedup~\cite{lowefficiency}. Moreover, some developers report that training may be stuck due to the drop-out of involved devices~\cite{dropout}.
These issues are quite essentially significant, as they not only affect the quality of DL models, but also incur a high cost of computation resources and developers' efforts. However, characterizing faults related to distributed training is missing. To the best of our knowledge, only Humbatova et al.~\cite{HumbatovaJBR0T20} mentioned a fault about data parallelism in distributed training. Zhang et al.~\cite{ZhangXZLLY20} characterized job failures in DL platforms, but did not discuss the specific features of distributed training. 

To fill in the knowledge gap, this paper presents the first comprehensive study on developers' issues in distributed training. 
Given the increasing dependence of current DL software on distributed training,
it is important to understand the relevant issues that developers encounter, so that researchers and framework vendors could help developers prevent, detect, and fix the common issues in a targeted manner. 
We aim to answer three research questions: 

\textbf{\revise{RQ1 (topics in how-to questions):}} \revise{What are the distributed training topics that developers frequently seek for help? To answer this question,} we investigate the common challenges that developers encounter in distributed training by analyzing the relevant how-to-questions, which indicate the distributed training knowledge that developers are inexpert at and thus tend to induce future faults.

\textbf{\revise{RQ2 (symptoms of faults):}} 
\revise{What are the fault symptoms that developers frequently encounter in distributed training? To answer this question,} we summarize the frequent software faults related to distributed training, which are overlooked by previous work, via constructing a comprehensive taxonomy of the fault symptoms.

\textbf{\revise{RQ3 (fix patterns):}} 
\revise{What are the common fix patterns for different fault symptoms in distribution training? To answer this question,} we study each symptom's common fix patterns to provide actionable insights for automated testing and repair techniques for distributed-training-related faults.

To collect the data of our interest, we focus our study on the three most popular DL frameworks, i.e., TensorFlow~\cite{AbadiBCCDDDGIIK16}, PyTorch~\cite{PaszkeGMLBCKLGA19}, and Keras~\cite{keras} that support distributed training and a widely-used DL framework that is specifically designed for distributed training, i.e., Horovod~\cite{abs-1802-05799}. 
Specifically, we construct a dataset of 1,131 distributed-training-related developers' issues that occur during the use of these frameworks from SO and GitHub, two commonly-used data sources for studying software issues~\cite{ZhangCCXZ18, FrancoGR17, HumbatovaJBR0T20, IslamNPR19, ChenCLW0L20}.

The results offer a series of findings that provide practical insights on better distributed training practice for developers, future research topics for researchers, and suggestions for DL framework vendors. We summarize the key findings and implications in Table \ref{tab:finding}. 
We make publicly available the code and the data in this study~\cite{dataset} as an additional contribution to the research community.


\section{Background}
\begin{figure}[t]
    \centering
    \includegraphics[width=0.6\linewidth]{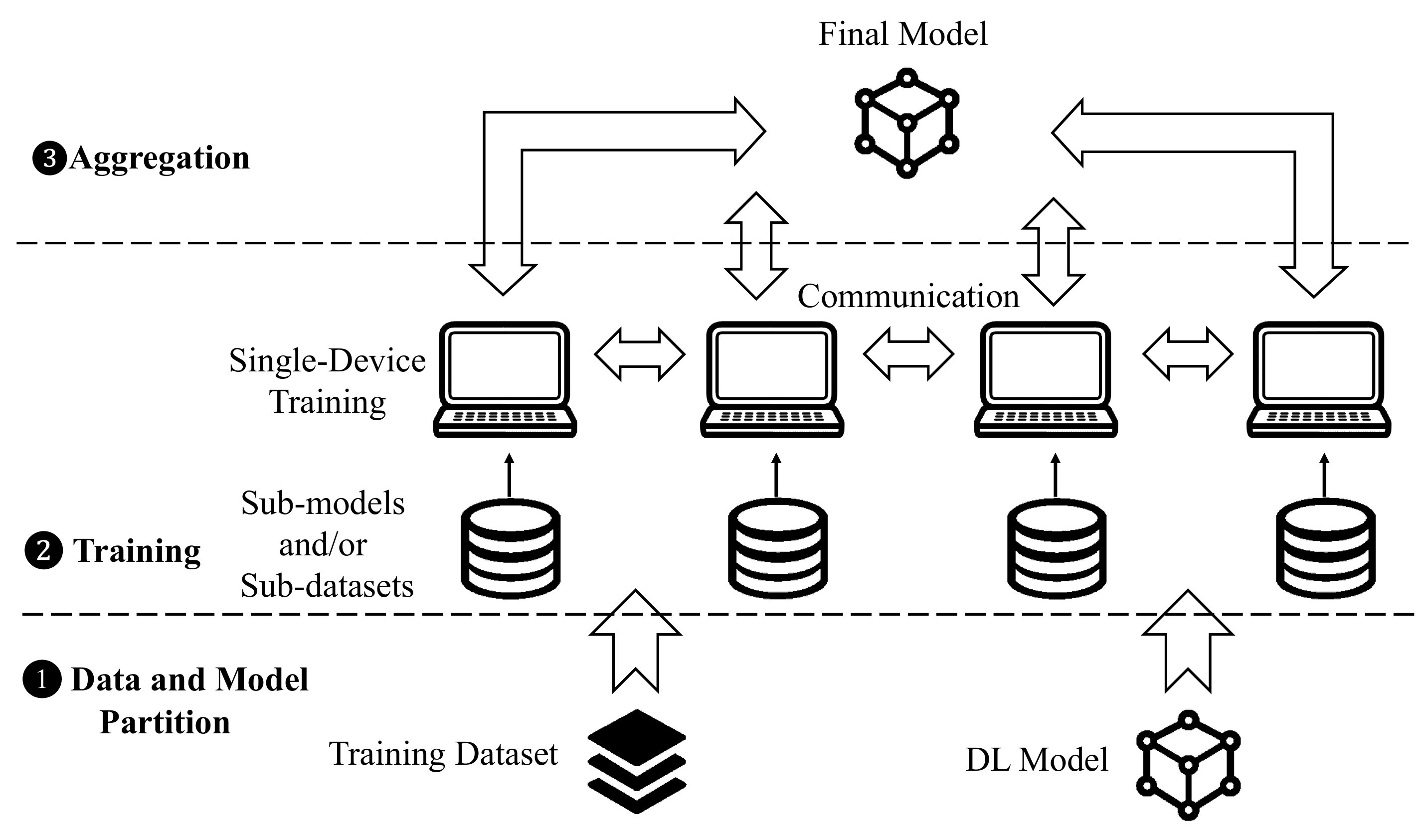}
        \vspace*{-3ex}
    \caption{Workflow of distributed training.}
    \label{fig:workflow}
\end{figure}
With the growing computation demands of DL, distributed training has been an important enabling technique for DL software. It parallelizes and distributes the computation and memory of DL training across multiple devices, e.g., GPUs, TPUs, and server machines. 
This process involves how to split training tasks, allocate computation resources, and coordinate various functional modules among different devices to achieve a balance between training speed and accuracy. 
To facilitate the understanding of distributed training, we present its common workflow in Fig.~\ref{fig:workflow}. 

\subsection{workflow}
A distributed training job first needs to be partitioned into multiple tasks that can run in parallel on different devices (\blackcircled{1}).
The most common parallelization ways are data parallelism and model parallelism~\cite{MayerJ20}.
For data parallelism, the training data is split into non-overlapping chunks, and then these data chunks are fed into different devices; each device loads an identical copy of the DL model~\cite{KrizhevskySH12};
For model parallelism, the DL model is split, and then each device loads a different part of the DL model for training~\cite{DeanCMCDLMRSTYN12}. 
Through data/model parallelism, the training data and the DL model are distributed on different devices.
Then, every device trains its own model with the data allocated to it (\blackcircled{2}). During this process, the devices communicate with each other to transfer essential data and synchronize the training progress on them. 
Finally, the trained models distributed on different devices are aggregated to obtain a new global model (\blackcircled{3}).
In the workflow, distributed training also relies on the environment, including hardware characteristics of devices, runtime environment (e.g., memory), network setting, and installed dependency libraries. 

\subsection{Related Work}
\noindent \textbf{Distributed training.}
With the increment of data size, model size, and computation requirement, distributed training has become a standard practice~\cite{JiaZA19}. 
Distributed training for DL comes with many possibilities for parallelization, among which data and model parallelism are predominant. 
In data parallelism\cite{KrizhevskySH12}, each worker (e.g., machines and GPUs) loads an identical copy of the DL model. Training data is split into non-overlapping chunks and fed into the model replicas of the workers. 
In model parallelism\cite{DeanCMCDLMRSTYN12}, the DL model is split, and each device loads a different part of the model. 
Apart from data and model parallelism, 
there are also novel parallelization methods such as hybrid~\cite{MayerJ20, JiaZA19} and pipeline parallelism\cite{HuangCBFCCLNLWC19}. 
With the innovation of parallelization methods, the distributed DL ecosystem has become rich and diverse~\cite{VerbraekenWKKVR20}.
DL frameworks such as TensorFlow~\cite{AbadiBCCDDDGIIK16} and  PyTorch\cite{PaszkeGMLBCKLGA19} support distributed training. Many distributed training frameworks and systems have also emerged, such as Horovod~\cite{abs-1802-05799},
BytePS~\cite{JiangZLYCG20}, PaddlePaddle~\cite{paddle}. 

\begin{table}[]
\caption{\revise{Summary of publication years, main objectives, and datasets in related work.}}
\small
\begin{tabular}{llll}
\hline
Paper                                    & Year & Main Objective   & Dataset                                          \\ \hline
\cite{ZhangCCXZ18}      & 2018   & \begin{tabular}[c]{@{}l@{}} Symptoms and root causes of TensorFlow program bugs and challenges \\in TensorFlow program bugs detection and localization.\end{tabular}                                           & SO posts, GitHub commits                      \\ \hline
\cite{IslamNPR19}       & 2019   & \begin{tabular}[c]{@{}l@{}} Types, root causes, impacts, prone stages, commonality, and evolution of \\bugs in the usage of DL libraries. \end{tabular}                              & SO posts, GitHub commits                      \\ \hline
\cite{ChenCLW0L20}      & 2020     & \begin{tabular}[c]{@{}l@{}}Popularity trend, difficulty, and taxonomy of challenges in deploying \\DL-based software.\end{tabular}                                                                       & SO posts                                     \\ \hline
\cite{IslamPNR20}       & 2020                                                           & \begin{tabular}[c]{@{}l@{}}Common bug fix patterns, fix pattern across bug types and libraries, risk in \\fix, and challenges of fixing bugs inside deep neural networks.\end{tabular} & SO posts, GitHub issues                   \\ \hline
\cite{HumbatovaJBR0T20} & 2020                                                                                         & The taxonomy of faults in DL systems.                                                                                                                                               & \begin{tabular}[c]{@{}l@{}}SO posts, GitHub issues, \\ developer interview\end{tabular} \\ \hline
\cite{ZhangXZLLY20}     & 2020                                                                                         & \begin{tabular}[c]{@{}l@{}}Failure error type, common root causes, and current testing and debugging \\practices in DL programming.\end{tabular}                       & \begin{tabular}[c]{@{}l@{}}Failed jobs in Philly, \\ developer interview\end{tabular}    \\ \hline
\cite{abs-2101-04930}   & 2021                                                        & Symptoms and fix strategies of DL-based mobile applications.                                                                                                                                                  & SO posts, GitHub issues                   \\ \hline
\end{tabular}
\label{tab:related}
\end{table}

\noindent \textbf{Empirical study on faults.} 
\revise{There have been a number of empirical studies that focus on faults in software systems, including traditional parallel computing systems and large-scale distributed systems~\cite{Asadollah18, GaoDQGW0HZW18,LiZLXLLX13}.}
\revise{However, distributed training is different from traditional parallel computing and traditional distributed programs in hardware devices they run on and program characteristics.} 
\revise{First, in parallel computing, the processors can access shared memory to share information between processors, whereas memory is usually not shared in distributed training~\cite{parallel_vs_distributed}. For example, a developer asked how to implement a sparse matrix in shared memory for parallel computing~\cite{so13239607}. This kind of issue rarely happens in distributed training because in distributed training the processors share information with communication between devices.}
\revise{Also, compared to traditional parallel computing and distributed computing, distributed training is more likely to run on GPU/TPU devices instead of CPU devices~\cite{distributed_training_device}. Therefore, the GPU-related and TPU-related faults (e.g., GPU device error) do not happen in traditional parallel computing or distributed programs.}
\revise{Moreover, many faults in traditional distributed programs are related to data processing and state consistency~\cite{GaoDQGW0HZW18}. 
These faults do not happen often in distributed training, because distributed training is compute-intensive instead of data-intensive, and many aggregation algorithms do not require strict consistency.
Therefore, existing studies on parallel computing and distributed program faults are not applicable to distributed training faults.}

With the rapid development of DL technologies, empirical studies on faults in software applications that make use of DL frameworks have emerged. 
\revise{Table~\ref{tab:related} summarizes the publication years, main objectives, and datasets of these empirical studies.}
\revise{Zhang et al.~\cite{ZhangCCXZ18} categorized four high level symptoms and seven root causes of TensorFlow program faults. They found that TensorFlow users relied on statistical values to determine test results and non-determinism was prevalent in the training process. Compared to our paper, this paper focused on only one DL framework. In addition, none of them focused on bugs related to distributed training or analyzed the fix patterns accordingly. }
\revise{Islam et al.~\cite{IslamNPR19} studied the characteristics of DL bugs. They found that data bugs and logic bugs were the most severe bug types in DL software and the major root causes of these bugs were incorrect model parameters and structural inefficiency. They characterized the bugs at a high level and did not focus on distributed training.}
\revise{Humbatova et al. ~\cite{HumbatovaJBR0T20} built a large taxonomy of faults in DL systems. Their taxonomy is mainly based on the root cause and include only one fault about data parallelism in distributed training.}
\revise{There are studies on the deployment challenges and faults of DL-based mobile applications ~\cite{abs-2101-04930, ChenCLW0L20}. Different from them, we focus on the training process instead of the deployment process and we focus on a specific domain, i.e., distributed training.}
Zhang et al.~\cite{ZhangXZLLY20} studied the symptoms, root causes, and fix patterns of job failures in a cloud-based DL platform. \revise{They found that 48.00\% of the failures occurred in the interaction with the platform rather than in the execution of code logic, mostly due to the discrepancies between local and platform execution environments and deep learning specific failures were mainly caused by inappropriate model parameters/structures and framework API misunderstanding. Distributed training jobs were included in the dataset, but they did not consider the differences between distributed jobs and single-device jobs when summarizing the failures. Our paper builds a fine-grained taxonomy of fault symptoms that includes distributed-specific symptoms and analyzes the fix patterns of distributed-specific faults.}

\subsection{Scope}
In this paper, we focus on the technical issues that developers encounter in distributed training. 
First, we analyze the frequent topics of general how-to questions about distributed training (\textbf{RQ1}). Then, we analyze the faults that occur during distributed training.
Specifically, we analyze the fault symptoms (\textbf{RQ2}) and distill common fix patterns for different symptoms (\textbf{RQ3}). 
Note that there are two kinds of faults during distributed training:  distributed-specific faults, which are caused by distributed-specific reasons (e.g., communication failure and invalid data partition), and non-distributed-specific faults, which are caused by non-distributed-specific reasons (e.g., wrong type of input data). 
Some of them share common symptoms (e.g., out of memory), although they are caused by different reasons. 
To show the whole picture of fault symptoms in distributed training, in RQ2, we construct our taxonomy based on both kinds of faults.
However, the fix patterns for non-distributed-specific faults have been comprehensively studied in previous work~\cite{ZhangCCXZ18, ZhangXZLLY20, IslamNPR19, IslamPNR20, HumbatovaJBR0T20}.
Therefore, in RQ3, we focus on only the fix patterns of distributed-specific ones.

\section{Methodology}
\label{methodology}
\begin{figure}[t]
    \centering
    \includegraphics[width=0.75\linewidth]{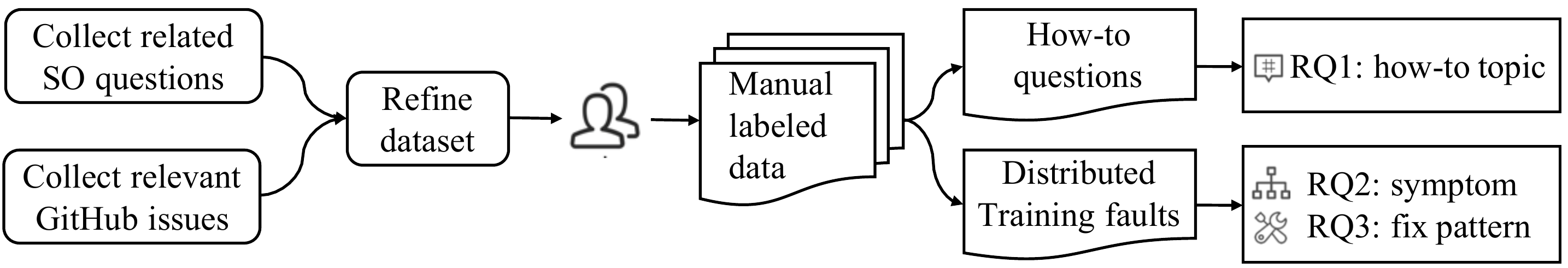}
    \caption{Overview of the methodology.}
    \label{fig:methodology}
\end{figure}

To characterize developers' issues in distributed training, we collect and analyze relevant SO questions and GitHub issues. The overview of our methodology is illustrated in Fig.~\ref{fig:methodology}. 

\subsection{Data Collection}
Since distributed training is mainly supported by state-of-the-art frameworks, we collect developers' issues that occur during the use of relevant frameworks to construct the dataset of our interest. Specifically, we focus our study on Horovod, which is the most popular framework specifically designed for distributed training and has been widely adopted in both academia~\cite{ZhangCLWAJ20, uber, nvidia2} and industry~\cite{uber, FdDL, databricks, nvidia, nvidia2}. In addition, we also consider the three most popular DL frameworks, i.e., TensorFlow, PyTorch, and Keras~\cite{frameworks1, frameworks2, frameworks3, frameworks4}, since all of them support distributed training. 

\subsubsection{Mining SO}
\label{sec:so}
SO is one of the most popular Q\&A websites where developers ask for help on unresolved technical issues~\cite{ChenCLW0L20}. It has been a commonly used data source for studying developers' software issues~\cite{ZhangCCXZ18, HumbatovaJBR0T20, IslamNPR19, ChenCLW0L20, IslamPNR20,sigsoftWenCLL00JL21}. 
Moreover, SO users range from novices to experts~\cite{ZhangCCXZ18}, increasing the diversity of collected issues. 

First, we download the entire SO dataset from the official Stack Exchange Data Dump~\cite{stackexchange} on December 6, 2021. The dataset (denoted as set $\mathcal{A}$) contains \revise{all of the questions that were ever posted on SO, covering a time period from July 31, 2008 to December 6, 2021}. Each question has one to five tags indicating its topics.
From $\mathcal{A}$, we extract 103,099 questions tagged with at least one of the four selected frameworks and denote these questions as set $\mathcal{B}$. 

\revise{SO questions in $\mathcal{B}$ are tagged with DL frameworks, but may contain noise that is not related to distributed training. For example, there are posts about traditional single-device DL~\cite{so57244733, so39863606}. 
Therefore, we need to further extract the distributed training-related questions from $\mathcal{B}$.} To this end, we randomly extract 1,000 questions from $\mathcal{B}$ and \revise{two authors discuss these questions carefully to manually identify a set of keywords that are highly related to distributed training.}
Next, we evaluate the recall level of these keywords, i.e., the percentage of the distributed-training-related posts that can be identified by these keywords. Specifically, we randomly select another 500 \revise{questions} to perform the evaluation and also identify new keywords from them. We add the new keywords to the keyword set after the evaluation. We repeat the above evaluation process four times until the keyword set can achieve a recall of 90\%. Note that here we do not consider the precision level of these keywords since any misidentified post can be filtered out during the refining process in Section \ref{refining} and will not threaten the validity of our results. As a result, we have the following keywords:
\{``\textit{distributed}'', 
``\textit{distribute}'', 
``\textit{parallel}'', 
``\textit{paralleled}'', 
``\textit{parallelism}'',
``\textit{data-parallel}'', ``\textit{dataparallel}'', ``\textit{model-parallel}'',
``\textit{modelparallel}'',
``\textit{workers}'',
``\textit{multi-server}'',
``\textit{multiple server}'', ``\textit{multiple servers}'',
``\textit{multi\_gpu}'', ``\textit{multi-gpus}'', ``\textit{multi-gpu}'', 
``\textit{multiple gpus}'', ``\textit{multiple gpu}'',
 ``\textit{multi-machine}'',
``\textit{multiple machines}'', ``\textit{multiple machine}''\}.
We perform a case-insensitive search within the title and body (excluding code snippets) of each question in $\mathcal{B}$ and identify 2,311 questions (denoted as set $\mathcal{C}$) that contain at least one of these keywords. \revise{The cumulative numbers of the questions in set $\mathcal{C}$ per year are shown in Fig.~\ref{fig:post}}.
Finally, we follow previous studies~\cite{HumbatovaJBR0T20, LouCCHZ20, abs-2101-04930} to exclude questions that do not have an accepted answer, ensuring that we consider only questions with a confirmed answer. 
As a result, we obtain a total of 724 questions from set $\mathcal{C}$ and denote them as set $\mathcal{D}$. 

\subsubsection{Mining GitHub}
\begin{table*}[]
\footnotesize
\caption{Rules to identify distributed-training-related issues on GitHub.}
\begin{tabular}{lllcr}
\hline
\textbf{Framework} & \textbf{Labels to identify distributed-training issues} & \textbf{Labels to exclude issues} & \textbf{\begin{tabular}[c]{@{}c@{}}Filter by \\ keywords\end{tabular}} & \textbf{\begin{tabular}[c]{@{}r@{}}\# extracted \\ GitHub issues\end{tabular}} \\ \hline
Horovod & N.A. & \begin{tabular}[c]{@{}l@{}}bug, enhancement, update docs, \\wontfix,  awaiting response\end{tabular} & \ding{56} & 762 \\ \hline
TensorFlow & comp:dist-strat & \begin{tabular}[c]{@{}l@{}}type:feature, type:bug, type:docs-bug, \\ stalled,  stat:awaiting response, \\type:docs-feature\end{tabular} & \ding{56} & 135 \\ \hline
PyTorch & \begin{tabular}[c]{@{}l@{}}oncall: distributed, module: ddp, \\module: multi-gpu,  module: data parallel, \\pt\_distributed\_rampup\end{tabular} & enhancement, feature, function request & \ding{56} & 726 \\ \hline
Keras & N.A. & \begin{tabular}[c]{@{}l@{}}type:bug/performance, type:feature, \\type:docs,  stale, Enhancement, \\stat:awaiting response\end{tabular} & \ding{52} & 305 \\ \hline
\end{tabular}
\label{tab:github}
\end{table*}

GitHub is another commonly used data source for studying software issues~\cite{FrancoGR17,abs-2101-04930,ZhangCCXZ18, HumbatovaJBR0T20}. In line with previous work~\cite{FrancoGR17,abs-2101-04930}, we mine issues posted in the official GitHub repositories (``GitHub issues'' for short) of the selected frameworks to identify developers' issues that occur during their use. 
Compared to commits, GitHub issues contain more information such as original reports and developers’ discussions~\cite{FrancoGR17}. 
This characteristic makes GitHub issues more suitable for studying the difficulties and faults encountered by developers.
In addition, on GitHub, framework vendors employ repository-specific keywords to label different types of GitHub issues, such as bug reports, feature requests, users' questions, etc. Following previous work~\cite{FrancoGR17, abs-2101-04930}, we leverage these labels of GitHub issues to help us identify relevant developers' issues. We use the GitHub search API~\cite{githubsearch} to extract these GitHub issues from the framework repositories on December 6, 2021. The detailed process is as follows.

For each framework, two authors jointly examine the labels in its GitHub repository to determine which labels are related to distributed training and then extract GitHub issues marked with these relevant labels. Since Horovod is specifically designed for distributed training, we take all of the GitHub issues in its repository into consideration no matter which labels they are marked. As for Keras, we cannot find any labels related to distributed training, so we use the keywords identified in Section \ref{sec:so} to extract relevant issues. 
Then, for each repository, we follow previous work~\cite{abs-2101-04930} to use labels to exclude GitHub issues about new feature requests, bugs in the framework itself, and problematic documents. 
Additionally, to ensure that we consider only closed issues with a confirmed solution, GitHub issues without answers or responses are excluded with the help of labels. 
The remaining issues are denoted as set $\mathcal{E}$. 
Table \ref{tab:github} shows the used labels and the number of identified GitHub issues in each repository, respectively.

\subsubsection{Refining dataset}\label{refining}
Two authors further manually examine all the extracted posts (i.e., SO questions in $\mathcal{C}$ and GitHub issues in $\mathcal{E}$) to refine the final dataset. Specifically, we jointly read each post and exclude posts that (1) do not have clear descriptions or solutions, (2) fix a bug in the framework itself rather than in distributed training program, or (3) are not related to distributed training.
The disagreements are all resolved with the involvement of an arbitrator, who has more than five years of experience in distributed training and has published many related papers in top-tier conferences.
\revise{For example, an author labelled a post that mentioned multiple GPU devices as an eligible post to be included in our study, while another author excluded this post because it did not mention model training~\cite{so42307975}. After analyzing the code snippets provided in the post and discussing it with the arbitrator, they reach an agreement that this post is about partitioning data to multiple GPUs and distributed training with the CIFAR10 dataset, which fits in the scope of our paper. Therefore, this post is included in our final dataset.} 
\revise{We measure the agreement during the data refining process with Cohen's Kappa ($\kappa$)~\cite{cohen1960coefficient}, which is commonly adopted for inter-rater agreement measurement~\cite{abs-2101-04930, IslamPNR20}. The $\kappa$ value is 0.95, indicating almost perfect agreement~\cite{landis1977measurement}.}
The final dataset for our study consists of 1,075 posts, including 511 posts about Horovod, 329 posts about TensorFlow, 157 posts about PyTorch, and 83 posts about Keras.\footnote{Note that an SO post may be tagged with multiple framework tags.}
The scale of our dataset is comparable and even larger than those used in existing fault-related empirical studies with manual labeling~\cite{ZhangCCXZ18, FrancoGR17, ChenCLW0L20, abs-2101-04930}. 
\subsection{Manual Labeling}
To distill how-to topics, symptoms, and fix patterns, we label every post in the refined dataset manually.
\revise{To get an overview of the entire dataset, we first conduct a pilot labeling procedure with 50\% of the dataset. We choose a 50\% dataset because it is sufficient for the authors to be familiar with the posts and the 50\% dataset left is also sufficient for validating the taxonomy for multiple rounds. During pilot labeling, we build a pilot taxonomy. Then, we validate the taxonomy built by pilot labeling with the rest of the dataset for five rounds and continuously refine the taxonomy.}
\revise{Specifically, we follow the procedure below. }
\subsubsection{Pilot labeling.}
First, we randomly sample 50\% of our dataset for pilot labeling. Two authors follow an open coding procedure~\cite{Seaman99} to summarize categories for how-to topics, symptoms, and fix patterns by jointly analyzing the sampled posts. Specifically, they read all the posts carefully to understand their context and assign each post with a set of labels describing (1) the how-to topic, which describes the how-to question briefly, (2) whether the fault is specific to distributed training, (3) the fault symptom, which shows what the fault looks like, and (4) the fix pattern, which tells how a fault is resolved. These labels are optional for each post. If a post is raising a how-to question (e.g., asking how to implement a specific distributed training task or inquiring conceptual knowledge about distributed training), it is labeled with only the how-to topic. Otherwise, a post with a clear fault description is labeled with whether it is distributed-specific, the fault symptom, and the fix pattern. 
Then, they construct taxonomies for how-to topics, symptoms, and fix patterns by grouping similar labels together into categories \revise{and finally
establish hierarchical taxonomies in a bottom-up way.}
The taxonomies are adjusted continuously in the construction process. 
A post is assigned to all related categories if it contains multiple how-to questions or faults.
During the pilot labeling process, any disagreement is resolved by the arbitrator mentioned before. All labels, categories, and taxonomies are approved by all participants.

\subsubsection{Reliability analysis.}
For reliability analysis, two authors independently label the remaining posts with how-to topics, whether they are distributed-specific, symptoms, and fix patterns based on the classification criteria generated in the pilot labeling. The posts that cannot be classified into the current taxonomies are labeled with a new category named \textit{Pending}. Specifically, the process of reliability analysis involves five rounds, each with 20\% of the remaining posts. In each round, we measure the inter-rater agreement of the independent labeling using Cohen’s Kappa ($\kappa$)~\cite{cohen1960coefficient}, which \revise{is suitable for the scenarios where two raters examine the same set of data and assign the data to a set of categories. In addition, we use fixed marginal kappa because we have a fixed set of categories that have been determined during the pilot study and fixed marginal kappa is suitable for such cases~\cite{brennan1981coefficient, randolph2005free}.}
It is also a widely-adopted metric in SE literature~\cite{abs-2101-04930, IslamPNR20}. After each round, with the help of the arbitrator, all the authors jointly solve the conflicts of labeling results and discuss the posts in \textit{Pending} category to
determine whether new categories need to be added. Then all the posts in \textit{Pending} are assigned to the adjusted taxonomies.


Table \ref{tab:label} reports the $\kappa$ values for the five rounds in reliability analysis. 
We also report the number of new categories added in each round. 
In the final round, no new category is added, indicating saturation for all categories; the $\kappa$ value is 0.88, indicating an almost perfect agreement~\cite{landis1977measurement}.

\begin{table}[]
\caption{The process of reliability analysis. The third column shows the number of newly added categories for how-to topics, symptoms, and fix patterns in each round, respectively.}
\begin{tabular}{crcc}
\hline
Round & \# analyzed posts & \# new categories & $\kappa$ \\ \hline
1 & 108 & 3/4/3 & 0.39 \\
2 & 108 & 0/5/8 & 0.66 \\
3 & 108 & 0/3/4 & 0.73 \\
4 & 108 & 0/1/2 & 0.78 \\
5 & 106 & 0/0/0 & 0.88 \\ \hline
\textbf{Total} & 538 & 3/13/17 & - \\ \hline
\end{tabular}
\label{tab:label}
\end{table}

In summary, among the 1,075 posts in pilot labeling and reliability analysis, we identify a total of 1,131 developers' issues, including 494 how-to questions and 637 faults. 
We merge the categories with few developers' issues (less than 1\% in how-to questions or less than 1\% in the faults of the corresponding stage in distributed training) together as the ``others'' category. Based on the 494 how-to questions, we answer RQ1 in Section \ref{rq1}; 
based on the 637 real-world faults, we answer RQ2 and RQ3 in Section \ref{rq2} and Section \ref{rq3}, respectively. 
\section{How-to topics (RQ1)}
\label{rq1}
\begin{figure}[t]
    \centering
    \includegraphics[width=0.65\linewidth]{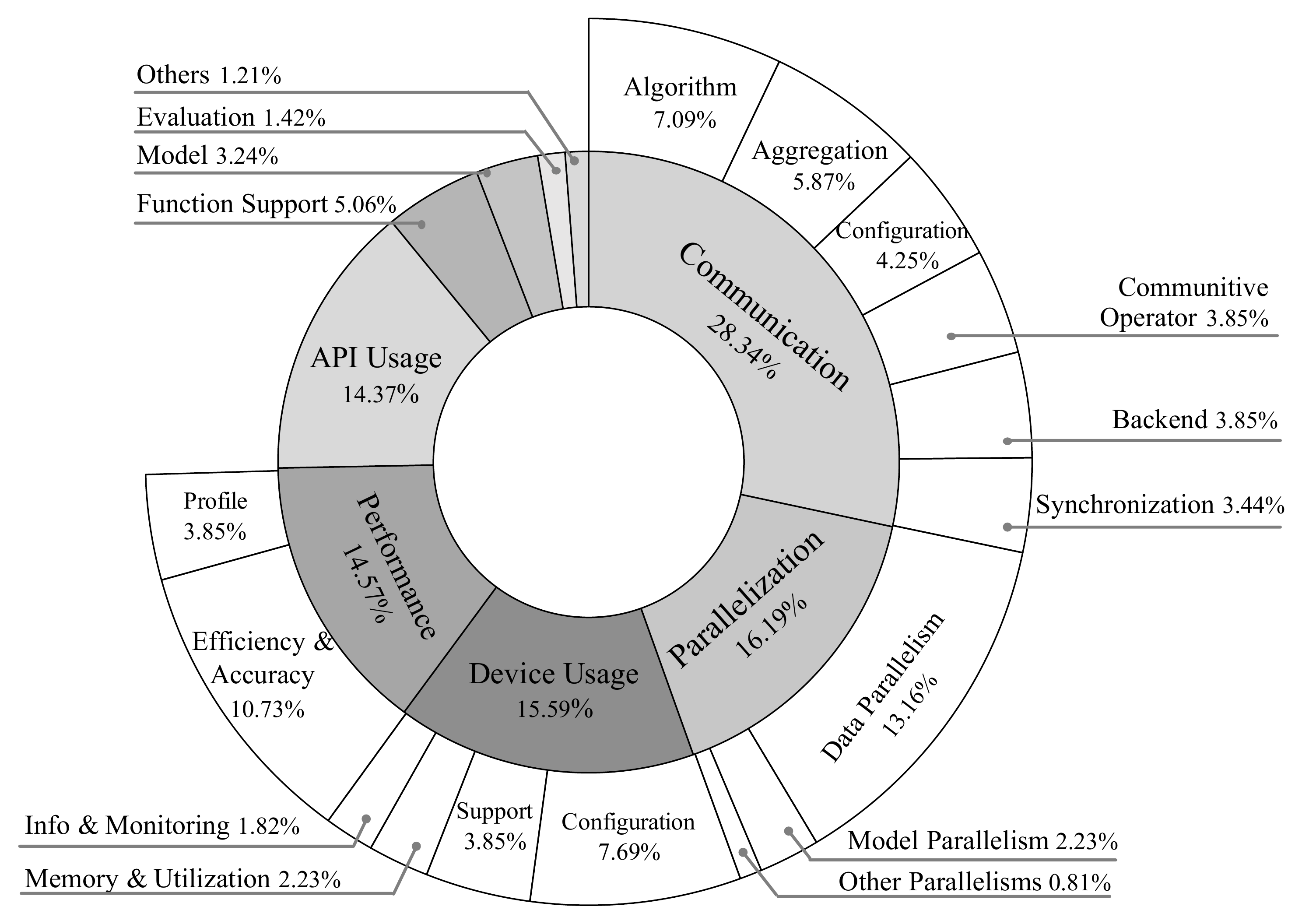}
    \caption{Topics in how-to questions of distributed training.}
    \label{fig:topics}
\end{figure}
Fig.~\ref{fig:topics} shows the hierarchical distribution of how-to topics in distributed training issues.
We observe that the topics asked by developers cover a wide spectrum of 9 high-level categories. \revise{There is no overlap between the topics in Fig.~\ref{fig:topics}, where communication, parallelization, device usage, model, and evaluation can be mapped to different steps in distributed training. The remaining three categories (i.e., performance, function support, and API usage) include general questions about distributed training and these questions are not about specific step(s) in distributed training. Questions about performance ask about the efficiency or accuracy of distributed training. Questions about function support are general ones about the current support of distributed training (e.g., whether distributed training frameworks support multi-GPU~\cite{keras106}). Questions about API usage ask about APIs that are related to distributed training but not specific to a step of distributed training. For example, a developer asked what changes to neural networks were made by the \texttt{strategy.scope()} API~\cite{so65358676}. This API covers most steps in distributed training, including data and model partition, data and model aggregation, and communication between devices, so it is not specific to any stage or component. Next, we will elaborate on the most frequent how-to topics.}

Communication is the most frequently asked topic (28.34\%).  
More specifically, 7.09\% of questions are related to algorithms about how to achieve the purpose of multi-device cooperative learning, such as ring all-reduce~\cite{ringallreduce} and parameter server~\cite{LiAPSAJLSS14}. 
5.87\% of the questions ask about data and model aggregation.
4.25\% of the questions are on the communication configuration. 
3.85\% of the questions are about collective communication operators (e.g., send, receive, broadcast, etc.). 
3.85\% are concerned about backend communication libraries such as NCCL~\cite{nccl} and gloo~\cite{gloo}.
The remaining questions about communication ask about synchronous or asynchronous training, such as the timing to update model parameters (i.e., weights and biases)~\cite{synchronization}.

The second most frequently asked (16.19\%) topic is parallelization, which describes how DL workflows are parallelized and run cooperatively. Most questions on parallelization are about the concept, support, or details of data parallelism (13.16\%). 
The rest 
are about model parallelism and other novel parallelization methods. 

15.59\% of how-to questions are about device usage. As multiple devices are involved in distributed training, configuring devices can be difficult; 7.69\% of questions are on device configuration. Developers also ask about the supported device usage of DL frameworks (e.g., whether Horovod supports training on multiple servers~\cite{devicesupport}). 
The rest are questions on memory usage, device utilization, device information, and monitoring. 

Developers are also concerned about
the performance of distributed training (14.57\%). 10.73\% are about the efficiency and accuracy of distributed training compared to non-distributed methods. Developers also ask about how to profile performance.


Overall, the how-to questions vary from naive concepts (e.g., basic knowledge) to very advanced algorithms (e.g., synchronization and aggregation), 
from general questions (e.g., training efficiency) to particular details (e.g., network setting). 
This diversity may owe to the huge differences between novices’ and experts’ posts on SO and GitHub, both of which reveal the difficulties and vulnerabilities in distributed training. 
\revise{We conduct Chi-Square test~\cite{greenwood1996guide} to compare the similarity of the how-to topic distribution for each of the two frameworks at a 95\% confidence level. The Chi-Square test is suitable for comparing the distribution of categorical data~\cite{greenwood1996guide} and suits our purpose well.} 
\revise{Since we carry out multiple tests, we adopt the Benjamini/Yekutieli method to adjust the p-values~\cite{ferreira2006benjamini}.}
\revise{We hypothesize that there is no significant difference between the observed frequencies of how-to topics across frameworks:}
\begin{align}
\revise{H_0: p_i^a = p_i^b, \forall~i~\leq~k},
\end{align}
\revise{where $a$ and $b$ are two frameworks and $k$ is the number of how-to topics. Table~\ref{tab:rq1} shows the adjusted p-values found from this test across the frameworks.}
We find that the adjusted p-values for all framework pairs are more than 0.05. This implies that the distributions of how-to topics in different frameworks are similar. 
\begin{table}[]
\caption{\revise{Adjusted p-values of the distributions of how-to topics between the frameworks}}
\begin{tabular}{|l|r|r|r|r|}
\hline
           & Horovod & TensorFlow & PyTorch & Keras \\ \hline
Horovod    & 1.000   & 0.897      & 1.000   & 0.897 \\ \hline
TensorFlow & 0.897   & 1.000      & 1.000   & 0.897 \\ \hline
PyTorch    & 1.000   & 1.000      & 1.000   & 1.000 \\ \hline
Keras      & 0.897   & 0.897      & 1.000   & 1.000 \\ \hline
\end{tabular}
\label{tab:rq1}
\end{table}

\textbf{For RQ1, see Findings F.1 and Implications I.1 in Table \ref{tab:finding}}.

\section{Symptoms (RQ2)}
\label{rq2}
\begin{figure*}[t]
    \centering
    \includegraphics[width=1\linewidth]{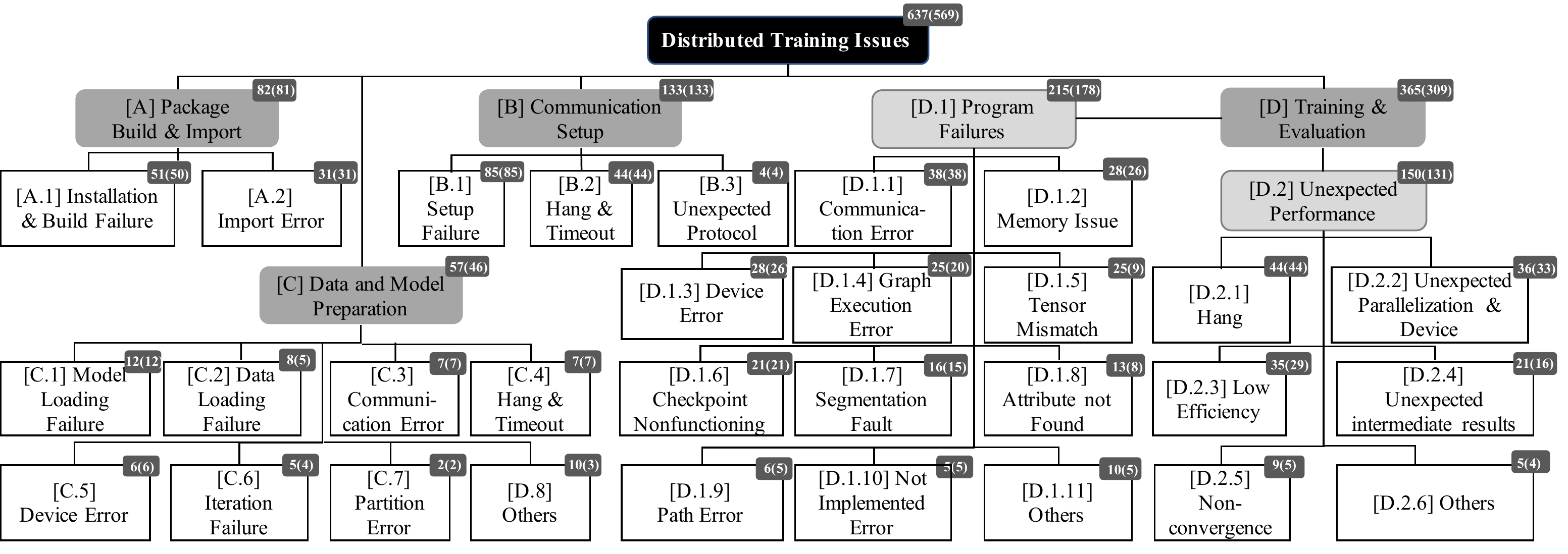}
        \vspace{-0.1in}
    \caption{Taxonomy of symptoms in distributed training. Each category has the number of corresponding faults in its top right corner with the number of distributed-specific faults in the bracket. }
    \label{fig:symptoms}
\end{figure*}
\revise{Some steps of distributed training have been integrated in the APIs of DL frameworks. For example, communication between devices and data/model aggregation are integrated in the \texttt{torch.nn.parallel.DistributedDataParallel} API in PyTorch. Therefore, we cannot tell the exact stage of distributed training that a fault belongs to from the symptom. Fortunately, we can accurately know whether a fault is caused by the used packages and whether a fault happens before training starts. Therefore, we construct the hierarchical taxonomy of fault symptoms in distributed training according to the programming steps of developers, which is shown in Fig.~\ref{fig:symptoms}.}
The root node consists of four children nodes, which are linked to four stages of distributed training. 
Each leaf node represents a category.
We use a number in the top right corner to represent the number of faults assigned to such a category and another number in brackets to represent the distributed-specific faults in this category. 


\textbf{Package build \& import (\textit{A}).} To write distributed training programs, developers 
import certain modules in their code (e.g., \textit{torch.distributed}) or build and install distributed training frameworks (e.g., Horovod). Faults that appear in this stage are included in the \textit{package build \& import} category, accounting for 12.87\% of the distributed training faults. 
62.20\% of faults in this stage happen when installing and building frameworks from source (i.e., \textit{installation \& build failure (A.1)}). 
Many developers reported that the error messages in this stage are difficult to understand~\cite{errormessage}. This makes it difficult for developers to resolve such faults and makes it difficult for us to further classify this category. 
Developers might also fail to import framework packages or certain package modules even though they have already installed frameworks successfully (i.e., \textit{import error (A.2)}). 

\textbf{Communication setup (B).}  \textit{Communication setup} is an essential step in distributed training when devices build up a topology for the communication in training. 20.88\% of faults related to distributed training show symptoms in this stage. As there is no need to set up communication in non-distributed training, all of the faults in this stage are specific to distributed training. 
63.91\% of faults in this stage are triggered when the devices cannot access each other correctly (\textit{setup failure (B.1)}). Besides, there are cases when the whole program is stuck at the communication setup stage or crashes because of timeout. We classify these issues into \textit{hang \& timeout (B.2)}. 
\textit{Unexpected protocol (B.3)} happens when devices do not communicate through the network protocol that developers set to. 

\textbf{Data and model preparation (C).} Before training, developers load or download training datasets; they also load or construct DL models to be trained. Then, as described in Fig.~\ref{fig:workflow}, developers split the datasets and models, and then distribute them to multiple devices for distributed training. Faults that appear in the above steps are included in the \textit{data and model preparation} category. We observe only a few related cases (8.95\%) in the entire dataset; 19.30\% of faults in this stage can also happen in non-distributed training. 
\textit{Model loading failure (C.1)} occurs when developers cannot load pre-trained models into memory. 
We use \textit{communication error (C.3)} to refer to program crashes because of unsuccessful communication between devices in this stage. 
The program might also hang or crash because of timeout (i.e., \textit{hang \& timeout (C.4)}) in this stage. 
\textit{Device error (C.5)} refers to program crashes because of invalid device assignments. 
Developers sometimes encounter problems with datasets (i.e., \textit{data loading failure (C.2)}, \textit{iteration failure (C.6)}, and \textit{partition error (C.7)}). 


\textbf{Training \& Evaluation (D).} \textit{Training \& evaluation (D)} is the most important stage of DL. 
It is also the largest category (57.30\% of identified faults) in our taxonomy, including a wide range of issues (17 symptom categories) related to all facets of training and evaluation. 
We classify these symptoms into two sub-categories: \textit{program failures (D.1)} and \textit{unexpected performance (D.2)}. 
\textit{Program failures (D.1)} refers to faults that lead to program crashes. \textit{Unexpected performance (D.2)} refers to cases when there is no crash but the programs do not behave as developers expect. 


There are various \textit{program failures (D.1)} symptoms in this stage. 
As common symptoms in both \textit{training \& evaluation} and \textit{data and model preparation}, \textit{communication error (D.1.1)} and \textit{device error (D.1.3)} account for 10.41\% and 7.67\% of faults in \textit{training \& evaluation}, respectively. 
7.67\% of the faults occur when there is illegal memory access or the memory is not enough for use (i.e., \textit{memory issue (D.1.2)}). 
\textit{Graph execution error (D.1.4)} occurs because of the improper computational graph that represents the architecture of the DL model, even though no symptom was shown before this stage. 
A few faults are caused by the shape or type of a tensor not matching its expectation (i.e., \textit{tensor mismatch (D.1.5)}), which is a common symptom in both distributed and non-distributed training. 
\textit{Checkpoint nonfunctioning (D.1.6)} is triggered when developers fail to save DL models. 
Sometimes, programs try to read or write an illegal memory location and trigger \textit{segmentation fault (D.1.7)}. Some faults are triggered due to reference to non-exist variables or functions (i.e., \textit{attribute not found (D.1.8)}). \textit{Path error (D.1.9)} refers to crashes because of unfound path references. \textit{Not Implemented Error (D.1.10)} happens when developers use functions or methods not implemented by frameworks. 
Even though some of these symptoms occur in non-distributed training as well, 
the root causes and fix patterns of them might be specific to distributed settings. We will discuss the details of root causes and fix patterns in Section \ref{rq3}. 

Some faults do not trigger failures explicitly, but generate problematic outputs or behave unexpectedly. 
As the most common symptom in this stage, 12.05\% of faults belong to \textit{hang (D.2.1)}, which means the program is stuck.
Sometimes the distributed training workflow does not parallelize on devices as expected. We classify these issues into \textit{unexpected parallelization \& device (D.2.2)}.
\textit{Low efficiency (D.2.3)} indicates distributed training does not achieve the expected speed-up. Such cases account for 9.59\% of faults identified in the stage. 
Developers also encounter faults when programs give problematic outputs (i.e., \textit{unexpected intermediate result (D.2.4)}) or the model does not converge (i.e., \textit{non-convergence (D.2.5)}). 

\revise{Compared to traditional single-device training, \textit{setup failure (B.1)}, \textit{unexpected protocol (B.3)}, \textit{partition error (C.7)}, and \textit{communication error (C.3, D.1.1)} are uniquely specific to distributed training and not been reported by previous studies~\cite{ZhangCCXZ18, ZhangXZLLY20, IslamNPR19, IslamPNR20, HumbatovaJBR0T20}. These symptoms are related to data partition or communication between workers, which are the specific steps of distributed training as shown in Fig.~\ref{fig:workflow}. }

\revise{For the rest of the symptoms, it is difficult to tell whether a fault is caused by distributed factors (e.g., data partition) or non-distributed factors (e.g., model architecture) from  only the symptom, which poses a big challenge to debugging. Developers can identify which component is responsible for a fault by troubleshooting, i.e., checking whether the distributed-specific modules work properly. Moreover, unit testing (i.e., testing one or more modules together) for distributed-specific modules can potentially help identify whether the fault is caused by the distributed-specific modules. For example, \textit{memory issue (D.1.2)} can be caused by setting a too-large batch size in the data loader, assigning data to the wrong devices, etc. To identify which component leads to the fault, developers can adopt unit testing on the relevant components
}

\textbf{For RQ2, see Findings F.2 and F.3, as well as Implications I.2 and I.3 in Table \ref{tab:finding}}. 

\section{Fix Patterns (RQ3)}
\label{rq3}
To capture how developers fix the observed distributed training faults, we summarize fix patterns for each symptom category. Since existing studies have already shown prevalent fix patterns for generic DL faults, here, we only focus on the fix patterns of the faults caused by distributed-specific mistakes. 
For the four stages in distributed training, we show the frequency of different fix patterns on their leaf categories in Fig.~\ref{fig:symptomA}, \ref{fig:symptomB}, \ref{fig:symptomC}, and \ref{fig:symptomD}. Due to space limit, patterns with low frequency (i.e., \#faults < 10 for \textit{training \& evaluation} and \#faults < 5 for other stages) are not shown. In each figure, the X-axis represents leaf categories with letter identifiers consistent with Fig.~\ref{fig:symptoms}; the Y-axis shows fix patterns following with their frequencies in the corresponding stage. We next elaborate on the prevalent fix patterns and demonstrate some real-world examples of faults and fixes. Except for the fix patterns that are already described, we present fix patterns for each stage in frequency order. 

\subsection{Faults in Package Build \& Import}
\label{sec:fixA}
\begin{figure}[t]
    \centering
    \includegraphics[width=0.55\linewidth]{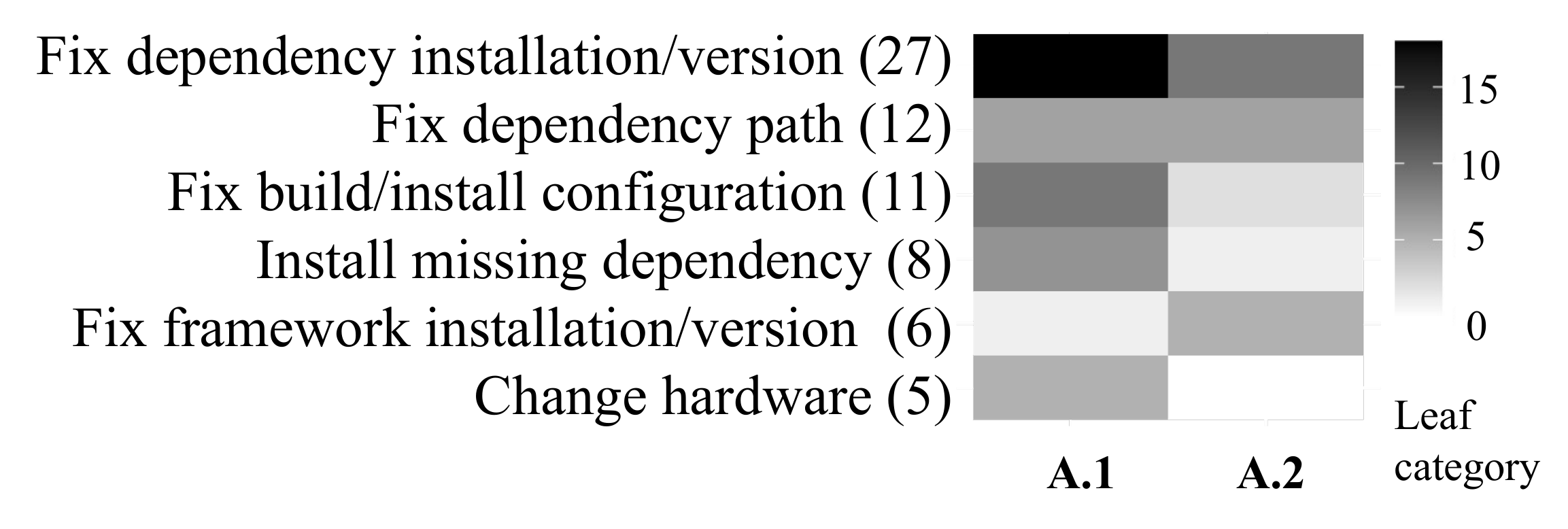}
        \vspace*{-3ex}
    \caption{Distribution of fix patterns for leaf categories in package build \& import issues.
    }
    \label{fig:symptomA}
\end{figure}

We identify six prevalent fix patterns for faults in \textit{package build \& import} and illustrate the distribution of these patterns on leaf categories in Fig.~\ref{fig:symptomA}. 

\textbf{Fix dependency installation/version \& install missing dependency.} 
43.21\% of distributed-specific faults in
\textit{package build \& import} are solved by re-installing DL framework dependencies, switching dependencies to a different version, or installing missing dependencies. 
These strategies are frequently adopted in both \textit{installation \& build failure (A.1)} and \textit{import error (A.2)}. The installation of DL frameworks usually relies on compilers and third-party libraries (such as communication libraries~\cite{gloo, nccl, mpi} and device-specific computing tools~\cite{cuda}). 
Horovod also relies on other DL frameworks (TensorFlow, PyTorch, etc.). Wrong installation or version of any dependency leads to failure in installation, build, or import. 
For example, a developer solved an installation failure by fixing CUDA version and NumPy installation~\cite{horovod161}). 

\textbf{Fix dependency path.} 
This pattern fixes 14.81\% of both \textit{installation \& build failure (A.1)} and \textit{import error (A.2)}. 
DL frameworks set default values for the paths where dependencies are installed. 
If developers install dependencies elsewhere, they should explicate the paths to dependencies in environmental variables. 
For example, a developer tried to install Horovod, but the build was unsuccessful because certain header files were not found~\cite {horovod1910}. She resolved the problem by adding header files of dependencies to the ``CPLUS\_INCLUDE\_PATH'' environmental variable. 

\textbf{Fix build/install configuration.} 
13.58\% of distributed-specific faults in \textit{package build \& import} are resolved by fixing the build or install configuration of DL frameworks, including fixing dependency library reference, fixing compilation options, etc. This fix pattern mainly resolves \textit{installation \& build failure (A.1)}. 


\textbf{Fix framework installation/version.} 
On one hand, some failures in building or installing frameworks are caused by broken framework packages or environment misconfiguration.
On the other hand, sometimes framework vendors fix bugs that lead to unsuccessful installation or build in the updated versions of frameworks. Therefore, re-installing DL frameworks or switching the frameworks to a different version fixes certain faults in \textit{package build \& import}. \revise{For example, in Example (a), a developer encountered an import error because the old versions of Horovod did not support the package she wanted to use~\cite{horovod818}. The fault was fixed by updating Horovod to a new version.}
\begin{figure}[t]
    \centering
\includegraphics[width=0.6\linewidth]{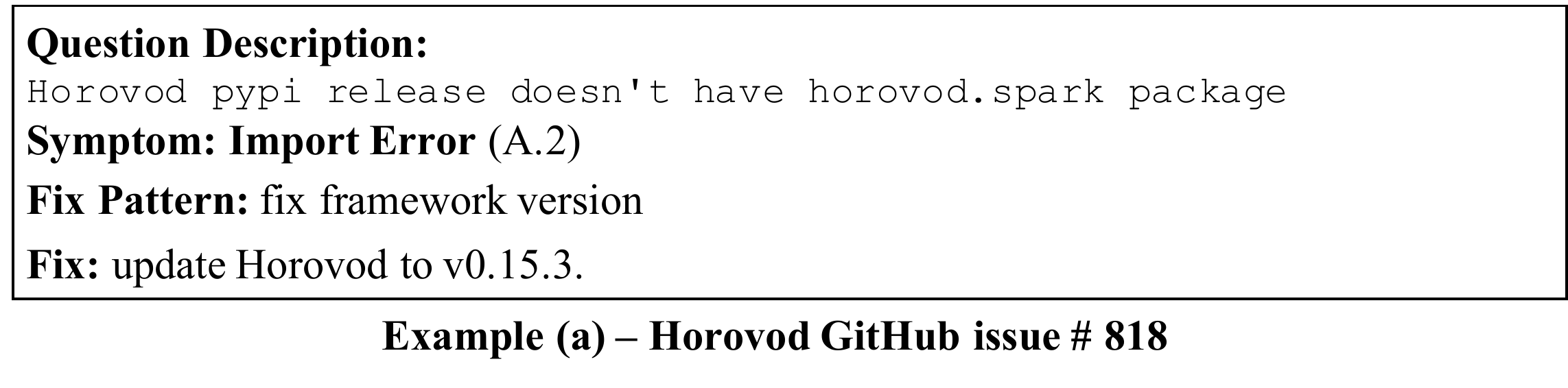}
    \label{fig:exampleA}
\end{figure}

\textbf{Change hardware.} 
Sometimes developers' hardware devices do not support the certain instruction set to build frameworks. In this case, the only approach to solving \textit{installation \& build failure (A.1)} is using devices with required supports. 
For example, a developer resolved an installation failure after switching to a server with CPUs that support AVX~\cite{horovod1798}). 

\revise{Most fixes in this stage are related to software dependencies, including framework installation, framework version, dependency installation, dependency version, dependency path, etc. Although many DL frameworks leverage dependency management tools such as Pip in installation, developers still need to fix the installation problems because the dependencies are too diverse and complex for these tools.}

\revise{\textbf{Please see Finding F.4 and Implication I.4 in Table~\ref{tab:finding} for fix patterns in \textit{package build \& import}.}}

\subsection{Faults in Communication Setup}
\label{sec:fixB}
\begin{figure}[t]
    \centering
    \includegraphics[width=0.6\linewidth]{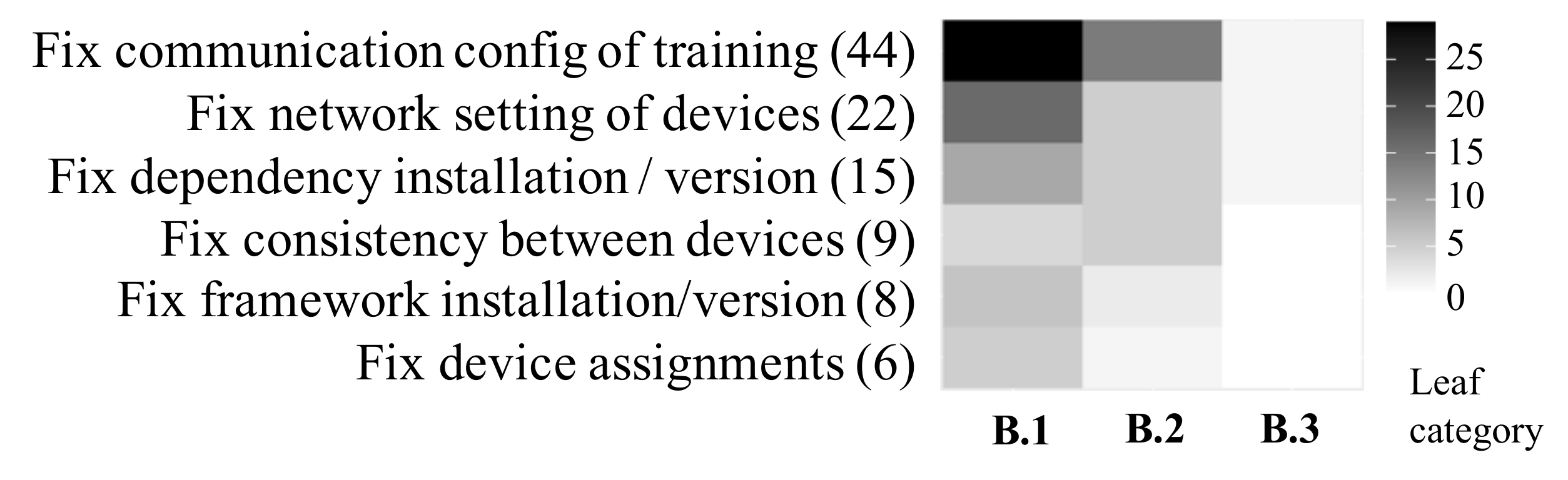}
    \caption{Distribution of fix patterns for leaf categories in communication setup issues.}
    \label{fig:symptomB}
\end{figure}

We identify six frequent fix patterns for faults in \textit{communication setup} and present the distribution of these patterns in Fig.~\ref{fig:symptomB}. 

\textbf{Fix communication configuration of training.}
Developers can configure the world size (i.e., the number of processes participating in communication), ranks (i.e., unique IDs of processes), and other configurations in the distributed settings.
Correctly configuring them mainly fixes \textit{setup failure (B.1)} and \textit{hang \& timeout (B.2)}, accounting for 33.08\% of distributed-specific faults in the stage. 
\revise{In Example (b), the developer encountered a fault when initializing RPC communication~\cite{pytorch54266}. The fix was to modify the configuration options.}

\begin{figure}[t]
    \centering
\includegraphics[width=0.6\linewidth]{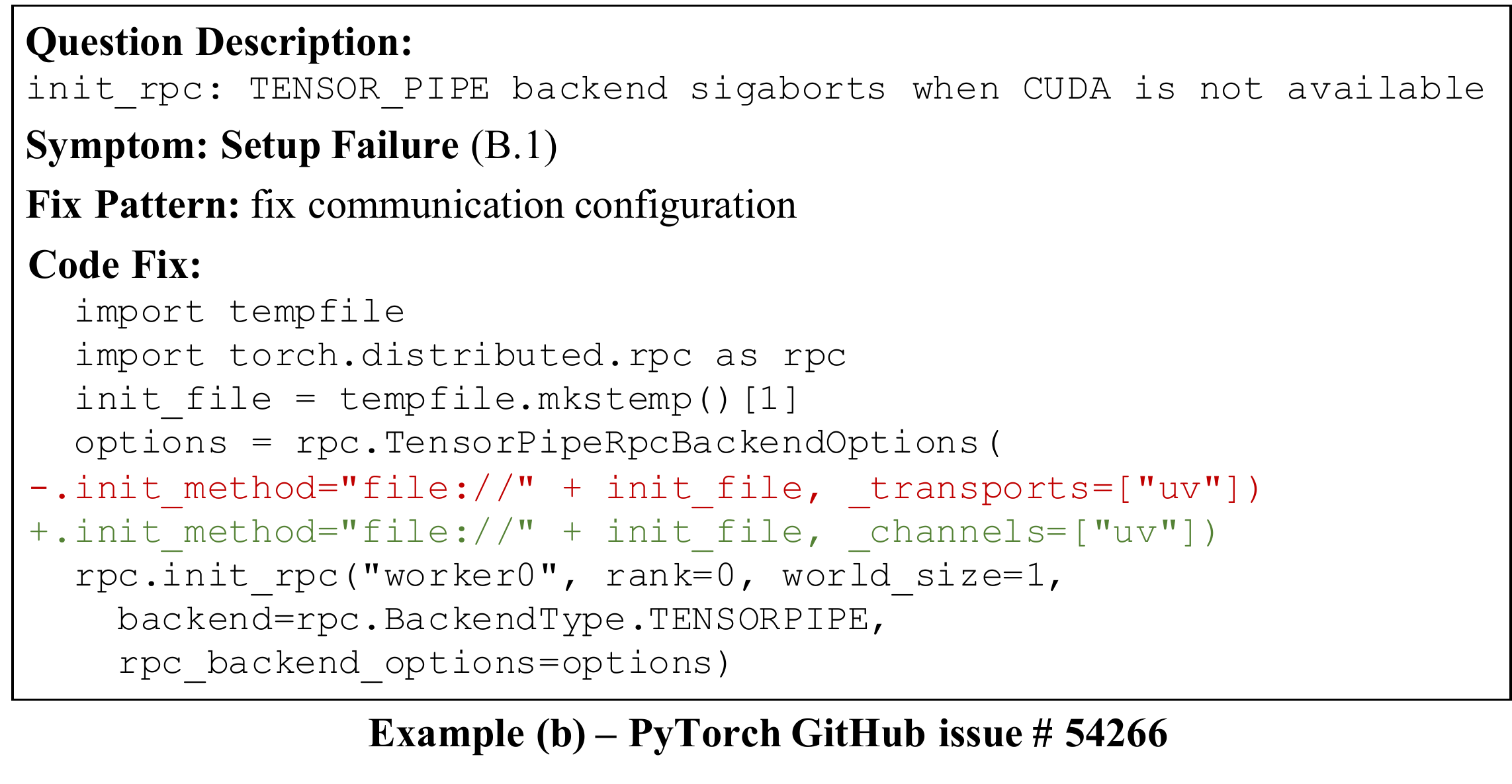}
    \label{fig:exampleB}
\end{figure}

\textbf{Fix network setting of devices.}
16.54\% of faults in \textit{communication setup} can be resolved by fixing network settings such as IP address, port, firewall, access permission, and so on. Wrong network setting is the main reason why devices cannot communicate with each other. The pattern can be adapted to all symptoms in this stage. For example, a developer could not build up the connection between two nodes because of ``permission denied''~\cite{horovod467}. The solution is to fix the public key setting for ssh. 

\textbf{Fix consistency between devices.} 
The inconsistency between different devices may lead to unsuccessful communication connections. 
For example, a developer had the problem of being unable to build up the communication connection between two nodes~\cite{horovod133}. She found out the reason was that the installation configurations of Open MPI on the two nodes were different. The final solution was to reinstall Open MPI with the same configuration on the nodes. 
Sometimes, if some devices are ready to train DL models whereas others are not, the inconsistency of device states attributes to unsuccessful communication setup as well. A developer reported that she got an error ``connection refused''~\cite{so38937984}. The reason was she did not successfully start training on the same number of devices as her topology configuration, leading to inconsistent device states. 

\textbf{Fix device assignments.} 
In \textit{communication setup}, each process should specify the device they are responsible for correctly, especially for the backends that rely on GPU-GPU communication (e.g., NCCL~\cite{nccl}). 

The remaining fix patterns have already been described in \S \ref{sec:fixA}. They are also applicable to faults in \textit{communication setup}.

\textbf{Fix dependency installation/version.}
Communication in distributed training depends largely on third-party communication libraries such as NCCL~\cite{nccl} and gloo~\cite{gloo}. 
The faults mainly belong to the symptom \textit{setup failure (B.1)} and \textit{hang \& timeout (B.2)}. 
For instance, a developer fixed the hang in Horovod when setting up communication by changing Open MPI version~\cite{horovod638}. 

\textbf{Fix framework installation/version.} 
This group of fixes re-install the DL framework or switch the framework to a different version. 
As DL frameworks and the distributed-related modules in DL frameworks are still in development, framework vendors fix bugs inside frameworks and update frameworks frequently. 
In addition, sometimes developers should change the framework version to make it compatible with dependencies.

\revise{In \textit{communication setup}, wrong communication configuration and wrong device network setting lead to most of the communication problems. The options of configurations and settings are diverse and scattered. This indicates that the communication configuration and network setting are too tedious to be set correctly. }

\revise{\textbf{Please see Finding F.5 and Implication I.5 in Table~\ref{tab:finding} for fix patterns in \textit{communication setup}.}}

\subsection{Faults in Data and Model Preparation}
\label{sec:fixC}
\begin{figure}[t]
    \centering
    \includegraphics[width=0.6\linewidth]{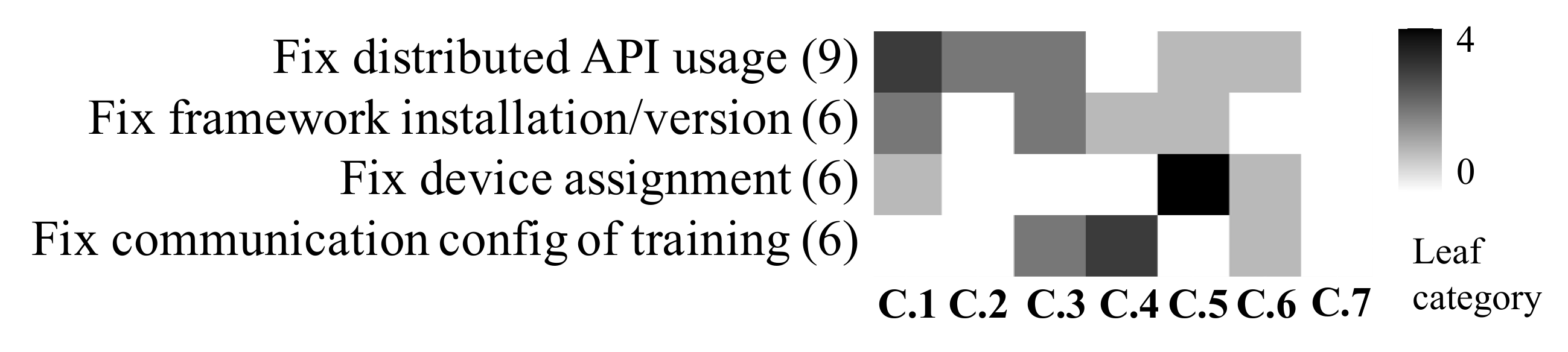}
    \caption{Distribution of fix patterns for leaf categories in data and model preparation issues.}
    \label{fig:symptomC}
\end{figure}

The solutions for distributed-specific faults in this stage are very diverse. Only four fix patterns are frequent.
These fix patterns are illustrated in Fig.~\ref{fig:symptomC}. 

\begin{figure}[t]
    \centering
    \includegraphics[width=0.8\linewidth]{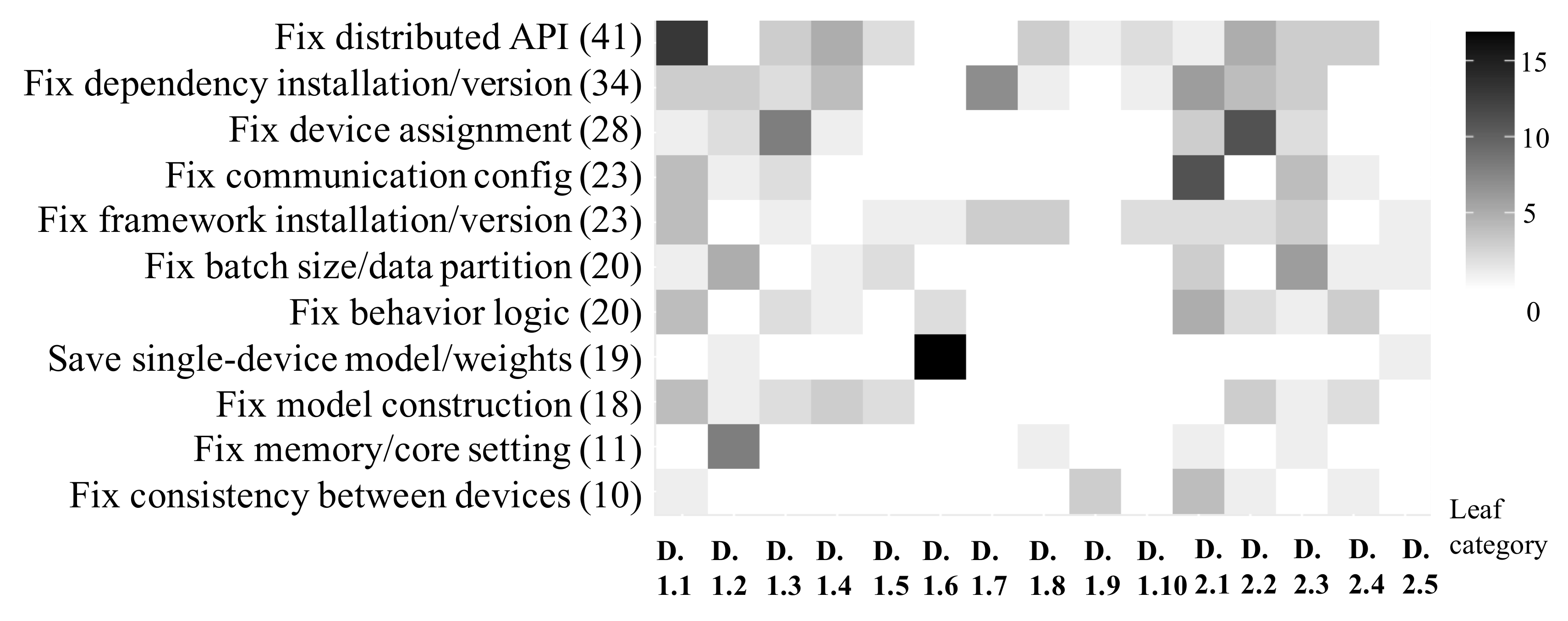}
    \caption{Distribution of fix patterns for leaf categories in training \& evaluation issues.}
    \label{fig:symptomD}
\end{figure}

\textbf{Fix distributed API usage.}
DL frameworks provide APIs for distributed training, such as  \texttt{torch.nn.DataParallel} and \texttt{torch.nn.parallel.DistributedDataParallel} in PyTorch and \texttt{tf.distribute.Strategy} in TensorFlow. Developers follow certain steps required by frameworks and write distributed training programs with these APIs. 
However, the complicated arguments of APIs and excessive procedures are difficult for developers to follow. For example, a developer could not initialize the communication topology because she forgot to call a certain API~\cite{pytorch38300}. The API is indispensable in distributed training with PyTorch. 

\textbf{Fix device assignment.}
In \textit{data and model preparation}, if data or model cannot be correctly assigned to corresponding devices, there will be a \textit{device error (C.5)}. Developers need to be careful with the device assignment (on type of device to use and the device id) of data and model to avoid \textit{device error (C.5)}.

The remaining fix patterns have been described in Section \ref{sec:fixA} and Section \ref{sec:fixB}. 

\textbf{Fix framework installation/version.}
Reinstalling framework or switching framework versions can also avoid faults in this stage, such as \textit{communication error (C.3)}, \textit{model loading failure (C.1)}, and so on. 
This is also because only certain framework versions provide mature support for some functionalities in distributed training. 

\textbf{Fix communication configuration of training.}
Communication is mandatory when assigning data and models to different devices. Fixing communication configuration helps avoid symptoms such as \textit{communication error (C.3)} and \textit{hang \& timeout (C.4)}. 

\revise{From the above fix patterns, we find that fix patterns such as fixing communication configuration and fixing framework installation/version are frequent in different stages of executing distributed training software. We also observe that there can be multiple fix patterns for one symptom, indicating that many fault symptoms in distributed training are
attributed to diverse factors. }

\revise{\textbf{Please see Finding F.6 and Implication I.6 in Table~\ref{tab:finding} for fix patterns in \textit{data and model preparation}.}}


\subsection{Faults in Training \& Evaluation}
\label{sec:fixD}
We identify 11 fix patterns for faults in \textit{training \& evaluation} stage, which includes the most symptoms and real-world faults. The distribution of these patterns is shown in Fig.~\ref{fig:symptomD}. 

\textbf{Fix batch size/data partition.}
This fix pattern mainly solves faults in \textit{memory issue (D.1.2)}, \textit{hang (D.2.1)}, and \textit{low efficiency (D.2.3)}. 
Batch size and data partition influence memory usage and distributed training efficiency. 
As distributed training introduces communication overheads and additional memory usage, only a proper batch size can make sure of high efficiency without out of memory faults. 
Besides, DL frameworks such as Horovod and Keras implement data parallelism naively. They require the dataset to be partitioned equally over devices. 
Otherwise, there might be a tensor shape mismatch problem or synchronization problem, because the number of data samples on different devices does not match up. 
For example, a developer encountered a tensor shape mismatch problem in distributed training~\cite{so43620478}. The solution was to make the number of samples divisible by $batch\_size \times N$, where $N$ is the number of GPU devices to use.
\begin{figure}[t]
    \centering
    \includegraphics[width=0.6\linewidth]{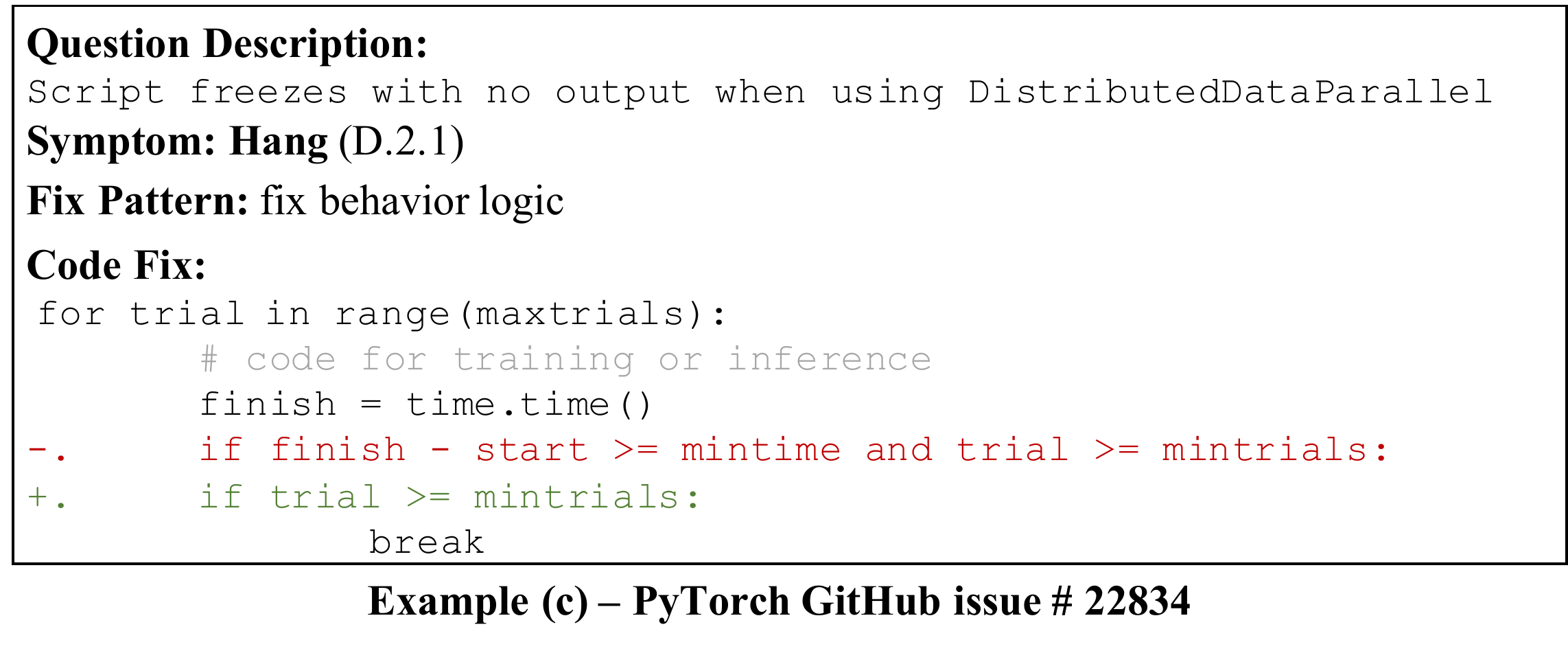}
    \label{fig:exampleC}
\end{figure}

\textbf{Fix behavior logic.}
Behavior logic refers to the logical relationship between the behaviors of different devices, such as profile writing and communication operations. 
Developers make mistakes in behavior logic when they are confused with the complicated logic or unfamiliar with distributed-related APIs. 
Inappropriate behavior logic leads to conflicts in distributed training or unexpected training performance. 
Example (c) shows a program hang problem~\cite{pytorch22834}. This is because the training speeds on different devices are not exactly the same, leading to one process exiting before the other one. To fix this fault, the developer needs to delete the timing code so that the training or inference on each device executes exactly the same number of steps.

\textbf{Save single-device model/weights only.} 
This pattern applies to only model saving problems (i.e., \textit{checkpoint nonfunctioning (D.1.6)}). In PyTorch and Keras, the ``single-device models'' and ``distributed-training models'' belong to different classes. In the case of unsuccessful model saving, saving the ``single-device model'' instead or saving model weights only is an effective workaround. 

\textbf{Fix model construction.}
Fixing how the model is constructed resolves faults in eight different symptoms. 
On one hand, model parallelism requires appropriate model partition. 
On the other hand, properly defining model layers and parameters (i.e., weights and biases) is essential for distributed training. For example, the symptom of the fault in Example (d) is \textit{device error (D.1.3)} which throws ``\textit{RuntimeError: arguments are located on different GPUs}''~\cite{so60799655}. This is because the developer did not define a certain tensor as an instance of \textit{torch.nn.Parameter} in her model. This results in the tensor 
not being assigned to certain GPU devices in graph replication. The corresponding solution is fixing the definition of this tensor in model construction. 
Although her model is not correctly constructed, such fault does not happen in non-distributed training as there is no need for graph replication. 
\begin{figure}[t]
    \centering
    \includegraphics[width=0.6\linewidth]{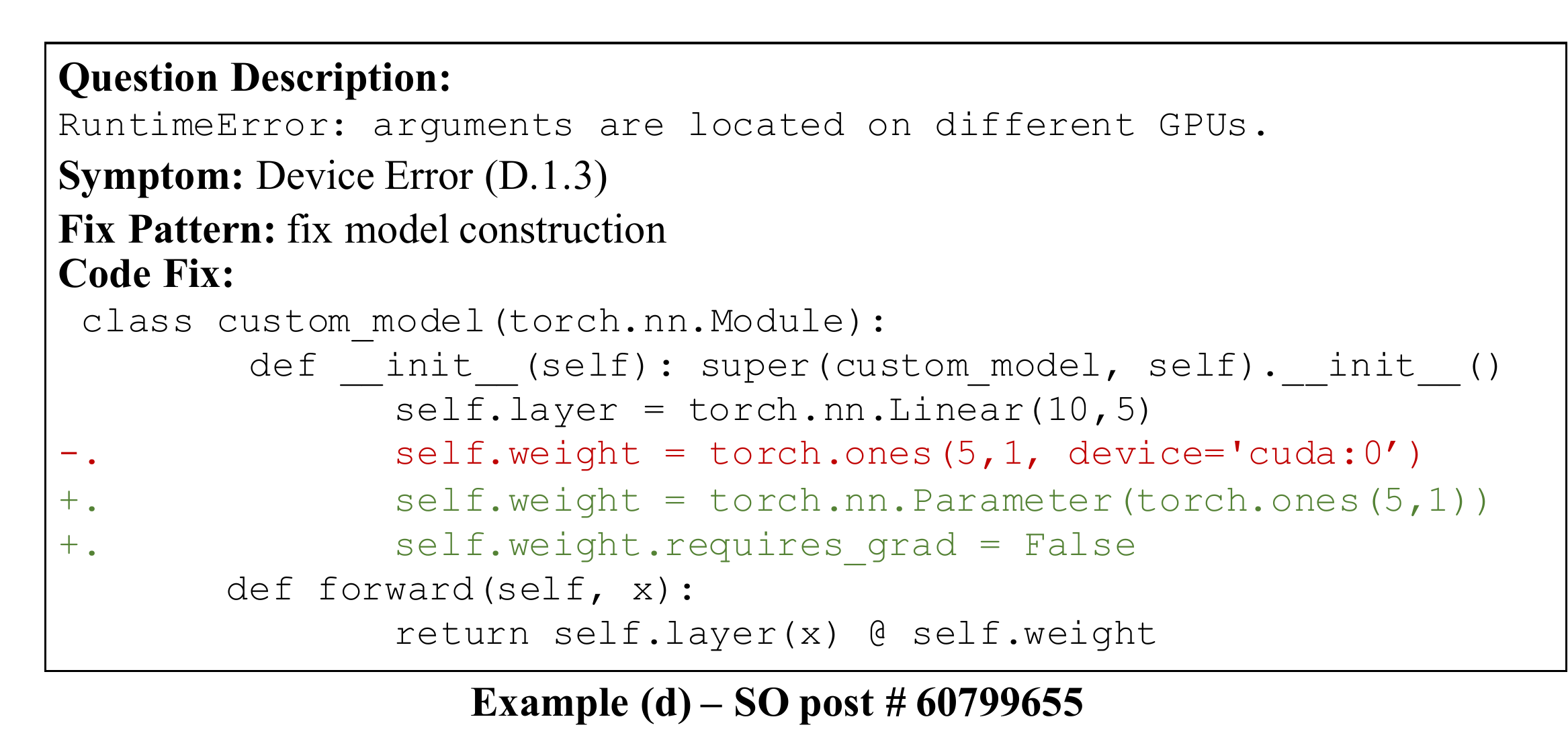}
    \label{fig:exampleD}
\end{figure}

\textbf{Fix memory/core setting.}
This group mainly resolves \textit{memory issue (D.1.2)} problems. By increasing the execution memory and cores in use, more resources will be allocated, which can resolve out of memory errors. Besides, modifying the configuration of how DL frameworks allocate memory is also effective~\cite{so45546737}. 




The remaining fix patterns have been described in Sections \ref{sec:fixA}, \ref{sec:fixB}, and \ref{sec:fixC}. They are also applicable to faults in \textit{training \& evaluation}. 

\textbf{Fix distributed API usage.}
This group fixes incorrect distributed API usage of developers or fixes hyperparameter configuration in these APIs. 
Since distributed APIs of frameworks control the whole distributed training procedure including data and model aggregation, synchronization, and so on, fixing distributed API usage resolves faults with almost every symptom in this stage.

\textbf{Fix dependency installation/version.}
Fixing dependency installation or version is the most frequent fix pattern in this stage. This strategy resolves 12.00\% of the distributed-specific faults with symptoms such as \textit{segmentation fault (D.1.7)} and \textit{hang (D.2.1)}.  

\textbf{Fix device assignment.}
Fixing device assignment of model or data mainly fixes \textit{device error (D.1.3)} and \textit{unexpected parallelization \& device (D.2.2)}.
For instance, a developer encountered such an unexpected parallelization behavior that TensorFlow allocates only one GPU device for computation~\cite{so43236349}. The fixing strategy is to modify the device allocation code and assign the model to every GPU device that is expected to be in use. 

\textbf{Fix communication configuration of training.} 
Wrong communication configurations lead to communication problems in \textit{training \& evaluation}. 
Therefore, fixing communication configuration mainly resolves \textit{communication error (D.1.1)} and \textit{hang (D.2.1)}. Some developers that encountered these faults also reported that they could not reproduce their faults~\cite{reporduce}. This is because multi-process communication
can easily cause nondeterministic behaviors, which makes fault reproduction difficult.

\textbf{Fix framework installation/version.}
This strategy also applies for \textit{training \& evaluation} stage.
On one hand, bugs in outdated frameworks may lead to \textit{segmentation fault (D.1.7)} symptoms. 
On the other hand, developers sometimes misuse APIs in a way unsupported by the current framework version, since APIs frequently evolve with DL frameworks. Therefore, developers should resolve such faults by changing the DL framework to a proper version. 
For example, a developer reported that she received ``\textit{RuntimeError: ProcessGroupNCCL does not support barrier}''~\cite{pytorch17848}. 
The corresponding fix is to upgrade PyTorch to v1.0.1 or a later version because ``barrier'' is not supported by the 1.0.x version. 

\textbf{Fix consistency between devices.} 
As we have described in Section \ref{sec:fixB}, consistent installation configurations and device states are essential for communication. These faults might not show until this stage. 
Besides, making sure that the code and datasets on all servers are in the exact same directory avoids \textit{path error (D.1.9)}. 


\revise{From all of the fix patterns of the 30 fault symptoms, we find that fixes related to communication are the most frequent, resolving faults in three out of the four stages.
Fixing dependency/framework installation/version, fixing distributed API usage, and fixing device assignment are also frequent; in total, they resolve up to 37.93\% of distributed-specific faults in distributed training, covering 25 frequent symptoms. Most of the faults in
20 out of the 30 symptoms can be fixed with no more than three fix
patterns, indicating that there are frequent fix patterns for these symptoms. We also found that about 47.25\%
of the fixes are system-level (including setting hardware devices, configuring environment, etc.) instead of training algorithm programming, among which 93.69\% are
distributed-specific. This indicates that compared to single-device training, the system-level configuration is challenging in distributed training. }

\revise{Although some symptoms also happen in traditional single-device training, the fix patterns of these faults can be different from the ones in single-device training. Many fix patterns of distributed-specific faults are to fix the modules that are specific to distributed training. These modules, such as process groups of communication and device assignment, are not required in traditional DL performed on a single device. For example,  for \textit{low efficiency (D.2.3)}, fixing batch size is a frequent fix pattern in single-device training, while fixing communication configuration and fixing data partition are frequent in distributed training.}

\revise{\textbf{Please see Finding F.7$\sim$9 and Implication I.6$\sim$8 in Table~\ref{tab:finding} for fix patterns in \textit{training \& evaluation} and the overall distributed training process.}}

\subsection{Fix Patterns across Frameworks}
To explore whether similar fix patterns occur on different frameworks, we study the distribution of fix patterns across frameworks. 
\revise{Similar to RQ1, we conduct Chi-Square test~\cite{greenwood1996guide} to compare the similarity of the fix pattern distribution for each of the two frameworks at a 95\% confidence level.} 
\revise{Since we carry out multiple tests, we adopt the Benjamini/Yekutieli method to adjust the p-values~\cite{ferreira2006benjamini}.}
\revise{We hypothesize that there is no significant difference between the observed frequencies of fix patterns across different frameworks:}
\begin{align}
\revise{H_0: p_j^a = p_j^b, \forall~j~\leq~q},
\end{align}
\revise{where $a$ and $b$ are two frameworks and $q$ is the number of fix patterns.} 
\revise{The adjusted p-value of framework pairs are shown in Table~\ref{tab:rq3}. Horovod-TensorFlow (0.052) and TensorFlow-PyTorch (1.000) are over 0.05}, which implies that the fix pattern distributions in these frameworks are statistically the same. This suggests that similar bug-fix patterns can be applied to different frameworks after being converted into a common intermediate representation.

\revise{\textbf{For fix patterns across frameworks, see Findings F.10 and Implications I.9 in Table \ref{tab:finding}}}. 
\begin{table}[]
\caption{\revise{Adjusted p-values of the distributions of fix patterns between the frameworks.}}
\begin{tabular}{|l|r|r|r|r|}
\hline
           & Horovod & TensorFlow & PyTorch & Keras \\ \hline
Horovod    & 1.000   & 0.052      & 0.022   & 7e-4  \\ \hline
TensorFlow & 0.052   & 1.000      & 1.000   & 0.002  \\ \hline
PyTorch    & 0.022   & 1.000      & 1.000   & 1e-4  \\ \hline
Keras      & 7e-4    & 0.002       & 1e-4    & 1.000 \\ \hline
\end{tabular}
\label{tab:rq3}
\end{table}

\begin{table*}[]
\caption{Summary of findings and implications.}
\vspace*{-1ex}
\footnotesize
\begin{tabular}{|L{6cm}|L{8cm}|}
\hline
\rowcolor{gray!50}
\textbf{Findings about how-to questions} &
  \textbf{Implications} \\ \hline
\textbf{F.1} Developers ask a wide range (9 high-level categories) of topics on distributed training. Communication, parallelization, device usage, and performance are most frequently asked\revise{, indicating that these topics are hot, challenging, and even confusing to developers.} &
  \textbf{I.1} 
  \revise{The frequency of performance-related questions indicates that current DL frameworks' implementations of distributed training are still far away from the expected performance in practice. For example, the  communication overhead has been evidenced to be a main cause of performance issues~\cite{JiangZLYCG20}. Therefore, we suggest DL framework vendors and researchers should carefully design useful communication optimization techniques. In practice, some techniques can be potentially promising. For example, the pipeline parallelism can better utilize all of the devices in distributed training~\cite{narayanan2019pipedream} and the gradient compression 
 can reduce the data to be transferred in communication~\cite{bai2021gradient}.}  \\ \hline
  \rowcolor{gray!50}
\textbf{Findings about faults symptoms} &
  \textbf{Implications} \\ \hline
\textbf{F.2} We construct a taxonomy of 30 fault symptoms for distributed training. Among them, there is no dominant symptom. The most common symptoms are hang, setup failure, and communication error, accounting for 35.32\% of the faults. &
  \textbf{I.2} The diverse and non-dominated symptoms suggest the challenge of designing automated tools for detecting and fixing distributed-training faults, which need to cover a broad spectrum of faults. \revise{The three most common symptoms are all related to communication. Researchers can pay more attention to these categories that developers find difficult to fix.} \\ \hline
\textbf{F.3} \revise{We find that 80.88\% of the non-distributed-specific faults show the same symptom as distributed-specific ones.} &
  \textbf{I.3} \revise{It is difficult to tell whether a fault is caused by distributed factors (e.g., data partition) or non-distributed factors (e.g., model architecture)  from  only the symptom, which poses a big challenge to debugging. Developers can identify which component is responsible for a fault by troubleshooting, i.e., checking whether the distributed-specific modules work adequately. Moreover, unit testing, i.e., testing one or more modules together, can potentially  help identify whether the fault is caused by the distributed-specific modules. For example, \textit{memory issue (D.1.2)} can be caused by setting a too-large batch size in the data loader, assigning data to the wrong devices, etc. To identify which component leads to the fault, developers can adopt unit testing on the relevant components.}\\ \hline
  \rowcolor{gray!50}
\textbf{Findings about fix patterns} &
  \textbf{Implications} \\ \hline
  \textbf{F.4} \revise{Most fixes in \textit{package build \& import} are related to the installation or version of frameworks and dependencies. This indicates that the currently-used dependency management tools such as Pip do not well support the complex and diverse dependencies of distributed training. } &
\textbf{I.4} \revise{Framework vendors can design dependency management and version management techniques that aim at the distributed environments to mitigate these problems (e.g., multi-device dependency check and automated version check). Developers can be more careful with package version requirements and can use virtual environments  or dockers  for package management. }\\ \hline
\textbf{F.5} \revise{In \textit{communication setup}, wrong communication configurations and wrong device network settings lead to most of the communication problems. The configuration and setting options are diverse and scattered. Not only are there many parameters in communication-related APIs, but also many environmental variables to be set. } &
\textbf{I.5} \revise{This finding indicates that the communication configurations and network settings are too tedious, error-prone, and time-consuming to be adequately set by developers, which heavily increases the  development cost. To help developers avoid and fix communication-related faults, the synthesis of software engineering and network systems can be worth exploring. For example, researchers can develop efficient testing and debugging techniques for communication  configuration, along with the synthesis of network configuration analysis~\cite{FogelFPWGMM15, Zheng:SIGCOMM22}.} \\ \hline
\textbf{F.6} Many faults in distributed training are attributed to diverse factors, indicating challenges in fault localization. For example, communication errors can be caused by misuse of distributed-training APIs, wrong dependency version,  wrong model construction, invalid network setting, etc. &
\multirow{2}{8cm}[2.6ex]{\textbf{I.6} \revise{ Framework vendors are encouraged to provide deeper hints for faults to assist developers’ resolution. For SE researchers, we suggest that they build runtime monitoring frameworks to collect traces for reproduction or adopt dynamic-analysis-based repair techniques. Existing fault reproduction methods such as checkpoint-and-replay may not be directly applied to distributed training because of the high runtime overhead or recovery overhead~\cite{WangLNMMTS19}. Researchers can design new multi-device checkpoint-and-replay techniques to help developers reproduce their faults efficiently.}} \\ 
\cline{1-1}
\textbf{F.7} Distributed training is usually multi-processing and can easily cause nondeterministic behaviors~\cite{WangLNMMTS19}. \revise{Sometimes developers cannot reproduce faults by running the same code again because of these characteristics of distributed training~\cite{reporduce}. This also makes fault localization challenging.} &
   \\ \hline
\textbf{F.8} For 20 out of the 30 symptoms, most of the issues in these categories can be fixed with no more than three fix patterns. &
\textbf{I.7} \revise{Our results of the frequent symptoms and their corresponding fix patterns also provide implications for testing of distributed training. When designing testing or debugging tools, researchers can  focus on the frequent and common fix patterns of these symptoms. For example, for \textit{communication error (C.3, D.1.1)}, developers can test the device assignments, device ranks, the IP address, the port, the world size, etc.; for \textit{tensor mismatch (D.1.5)}, the batch size, parameters of data partition, and parameters of model layers can be tested to locate the fault. For symptoms without frequent fix patterns, there is still a lot of space and challenges to integrate more existing patterns and explore more fix strategies.
} \\ \hline
\textbf{F.9} About 52.75\% of the issues can be resolved through programming. The fix patterns of the remaining 47.25\% are related to hardware devices, environment, and configurations, among which 93.69\% are distributed-specific. &
\textbf{I.8} \revise{
The system-level configurations are challenging for developers who do not have expertise knowledge of underlying systems. To alleviate low-level device management and environment configurations for developers, one solution is to alleviate these efforts in a serverless way, i.e., cloud providers provide efficient APIs by abstracting the low-level and tedious system configurations and allocate resources on demand according to the training jobs~\cite{jonas2019cloud, Gu:AsPLOS23}. This has been demonstrated to be effective for reducing developers' programming efforts~\cite{Liu0ZMB14}} \\ \hline
\textbf{F.10} \revise{Horovod, Tensorflow, and PyTorch show similar fix pattern distribution.} &
\textbf{I.9} Similar automatic bug fix tools may be reused for these frameworks after being converted into a common intermediate representation. \\ \hline
\end{tabular}
\label{tab:finding}
\end{table*}

\section{Threats to Validity}

\textbf{Selection of frameworks, keywords, and labels.}
First of all, the selection of frameworks may lead to possible selection bias in this study.
To mitigate this threat, we focus on the three most commonly-used DL frameworks and Horovod, which is widely adopted for distributed training. 
In addition, the keyword- and label-matching identification may result in false positive posts and loss of relevant posts.
The false positives are all discarded during the refining process in Section \ref{refining}. 
Moreover, as mentioned in Section \ref{sec:so}, our keywords have a high level of recall (i.e., 90\%), ensuring that most of the relevant issues can be identified. As for the label-matching identification,
\revise{to reduce manual efforts, we first follow previous work~\cite{FrancoGR17,abs-2101-04930} to automatically filter the collected dataset by a set of rules, which may result in a potential threat to the validity. For example, we directly exclude new feature requests and reports about bugs in frameworks from GitHub. However, there could be data in them related to  facilitating distributed training.}

\noindent \textbf{Selection of data sources.} 
\revise{Given the rapid evolution of distributed training, it is possible that new challenges and bugs can emerge in the future. Following the same methodology presented in this paper, we can reproduce the study with new industry and academia efforts of distributed training, in order to keep our results updated. Also,} it is impossible to collect all the issues about distributed training of DL software in the world, which may lead to a threat to the external validity of our study. To mitigate this threat, we select SO and GitHub, the two most widely-used data sources in empirical studies in the SE community~\cite{ZhangCCXZ18, IslamNPR19, HumbatovaJBR0T20, AghajaniNVLMBL19, IslamPNR20}, to collect representative real-world issues reported by developers. 

\noindent \textbf{Subjectivity of researchers.} The subjectivity in manual labeling presents a possible internal threat to the validity of our results. 
To minimize this threat, we follow the widely-adopted open coding procedure, in which two authors are involved in inspecting cases and another experienced arbitrator helps to reach an agreement through discussions. 
We also use Cohen's Kappa to measure the inter-rater agreement of independent labeling. The high kappa values indicate almost perfect inter-rater agreement.

\section{Conclusion}
In this paper, we presented an empirical study on issues in distributed training of DL software by manually inspecting 1,131 related issues from Stack Overflow and GitHub. We distilled frequent topics in developers' how-to questions. 
We also constructed a fine-granularity taxonomy of 30 fault symptom categories and summarized fix patterns for different fault symptoms. 
Our findings are helpful to developers of distributed training software and the framework vendors of distributed training platforms.
In the future, researchers can develop debugging, testing, and auto-configuration tools based on the frequent combinations of fault symptoms and fix patterns, our findings, and insights. 


\begin{acks}
This work was supported by the National Natural Science Foundation of China under the grant numbers 62172008 and 62102009, and the National Natural Science Fund for the Excellent Young Scientists Fund Program (Overseas).
Zhenpeng Chen was supported by the ERC Advanced Grant under the grant number 741278 (EPIC: Evolutionary Program Improvement Collaborators). Zhenpeng Chen and Xin Jin are corresponding authors of this paper.

\end{acks}

\bibliographystyle{ACM-Reference-Format}
\bibliography{paper}


\begin{thebibliography}{105}


\ifx \showCODEN    \undefined \def \showCODEN     #1{\unskip}     \fi
\ifx \showDOI      \undefined \def \showDOI       #1{#1}\fi
\ifx \showISBNx    \undefined \def \showISBNx     #1{\unskip}     \fi
\ifx \showISBNxiii \undefined \def \showISBNxiii  #1{\unskip}     \fi
\ifx \showISSN     \undefined \def \showISSN      #1{\unskip}     \fi
\ifx \showLCCN     \undefined \def \showLCCN      #1{\unskip}     \fi
\ifx \shownote     \undefined \def \shownote      #1{#1}          \fi
\ifx \showarticletitle \undefined \def \showarticletitle #1{#1}   \fi
\ifx \showURL      \undefined \def \showURL       {\relax}        \fi
\providecommand\bibfield[2]{#2}
\providecommand\bibinfo[2]{#2}
\providecommand\natexlab[1]{#1}
\providecommand\showeprint[2][]{arXiv:#2}

\bibitem[so1(2012)]%
        {so13239607}
 \bibinfo{year}{2012}\natexlab{}.
\newblock \bibinfo{title}{Parallel array or array of structures [closed]}.
\newblock
  \bibinfo{howpublished}{\url{https://stackoverflow.com/questions/13239607}}.
\newblock
\newblock
\shownote{Retrieved on December 21, 2022}.


\bibitem[so3(2016a)]%
        {so38937984}
 \bibinfo{year}{2016}\natexlab{a}.
\newblock \bibinfo{title}{{Distributed tensorflow on localhosts failed by
  ``socket error, connection refused''}}.
\newblock
  \bibinfo{howpublished}{\url{https://stackoverflow.com/questions/38937984}}.
\newblock
\newblock
\shownote{Retrieved on March 16, 2022}.


\bibitem[syn(2016)]%
        {synchronization}
 \bibinfo{year}{2016}\natexlab{}.
\newblock \bibinfo{title}{Synchronous vs asynchronous computation in
  Tensorflow}.
\newblock
  \bibinfo{howpublished}{\url{https://stackoverflow.com/questions/34349316/synchronous-vs-asynchronous-computation-in-tensorflow}}.
\newblock
\newblock
\shownote{Retrieved on March 16, 2022}.


\bibitem[so3(2016b)]%
        {so39863606}
 \bibinfo{year}{2016}\natexlab{b}.
\newblock \bibinfo{title}{Why neural network tends to output 'mean value'?}
\newblock
  \bibinfo{howpublished}{\url{https://stackoverflow.com/questions/39863606}}.
\newblock
\newblock
\shownote{Retrieved on December 21, 2022}.


\bibitem[rin(2017)]%
        {ringallreduce}
 \bibinfo{year}{2017}\natexlab{}.
\newblock \bibinfo{title}{Baidu-Allreduce}.
\newblock
  \bibinfo{howpublished}{\url{https://github.com/baidu-research/baidu-allreduce}}.
\newblock
\newblock
\shownote{Retrieved on March 16, 2022}.


\bibitem[so4(2017a)]%
        {so45546737}
 \bibinfo{year}{2017}\natexlab{a}.
\newblock \bibinfo{title}{CUDA\_ERROR\_OUT\_OF\_MEMORY: How to activate
  multiple GPUs from Keras in Tensorflow}.
\newblock
  \bibinfo{howpublished}{\url{https://stackoverflow.com/questions/45546737}}.
\newblock
\newblock
\shownote{Retrieved on March 16, 2022}.


\bibitem[dev(2017)]%
        {devicesupport}
 \bibinfo{year}{2017}\natexlab{}.
\newblock \bibinfo{title}{Horovod's Work Pattern?}
\newblock
  \bibinfo{howpublished}{\url{https://github.com/horovod/horovod/issues/117}}.
\newblock
\newblock
\shownote{Retrieved on March 16, 2022}.


\bibitem[so4(2017b)]%
        {so43236349}
 \bibinfo{year}{2017}\natexlab{b}.
\newblock \bibinfo{title}{How to use multiple GPUs effectively when training
  deep networks?}
\newblock
  \bibinfo{howpublished}{\url{https://stackoverflow.com/questions/43236349}}.
\newblock
\newblock
\shownote{Retrieved on March 16, 2022}.


\bibitem[so4(2017c)]%
        {so43620478}
 \bibinfo{year}{2017}\natexlab{c}.
\newblock \bibinfo{title}{Keras predict not working for multiple GPU's}.
\newblock
  \bibinfo{howpublished}{\url{https://stackoverflow.com/questions/43620478}}.
\newblock
\newblock
\shownote{Retrieved on March 16, 2022}.


\bibitem[so4(2017d)]%
        {so42307975}
 \bibinfo{year}{2017}\natexlab{d}.
\newblock \bibinfo{title}{Memory management when using GPU in TensorFlow?}
\newblock
  \bibinfo{howpublished}{\url{https://stackoverflow.com/questions/42307975}}.
\newblock
\newblock
\shownote{Retrieved on December 21, 2022}.


\bibitem[hor(2017)]%
        {horovod133}
 \bibinfo{year}{2017}\natexlab{}.
\newblock \bibinfo{title}{{Run distributed : ERROR: ORTE\_ERROR\_LOG: Data
  unpack would read past end of buffer}}.
\newblock
  \bibinfo{howpublished}{\url{https://github.com/horovod/horovod/issues/133}}.
\newblock
\newblock
\shownote{Retrieved on March 16, 2022}.


\bibitem[com(2017)]%
        {commconfig}
 \bibinfo{year}{2017}\natexlab{}.
\newblock \bibinfo{title}{Tensorflow: is There a Rule to Set the Port of
  Worker/PS When Creating ClusterSpec?}
\newblock
  \bibinfo{howpublished}{\url{https://stackoverflow.com/questions/41649708/tensorflow-is-there-a-rule-to-set-the-port-of-worker-ps-when-creating-clustersp}}.
\newblock
\newblock
\shownote{Retrieved on March 16, 2022}.


\bibitem[par(2018)]%
        {parallel_vs_distributed}
 \bibinfo{year}{2018}\natexlab{}.
\newblock \bibinfo{title}{Difference Between Parallel and Distributed
  Computing}.
\newblock
  \bibinfo{howpublished}{\url{https://www.differencebetween.com/difference-between-parallel-and-vs-distributed-computing/}}.
\newblock
\newblock
\shownote{Retrieved on April 18, 2023}.


\bibitem[hor(2018a)]%
        {horovod161}
 \bibinfo{year}{2018}\natexlab{a}.
\newblock \bibinfo{title}{{Horovod dosn't work with CUDA 9.1}}.
\newblock
  \bibinfo{howpublished}{\url{https://github.com/horovod/horovod/issues/161}}.
\newblock
\newblock
\shownote{Retrieved on March 16, 2022}.


\bibitem[hor(2018b)]%
        {horovod638}
 \bibinfo{year}{2018}\natexlab{b}.
\newblock \bibinfo{title}{Horovod hangs with multi gpus on one machine}.
\newblock
  \bibinfo{howpublished}{\url{https://github.com/horovod/horovod/issues/638}}.
\newblock
\newblock
\shownote{Retrieved on March 16, 2022}.


\bibitem[dat(2018)]%
        {databricks}
 \bibinfo{year}{2018}\natexlab{}.
\newblock \bibinfo{title}{Introducing HorovodRunner for Distributed Deep
  Learning Training}.
\newblock
  \bibinfo{howpublished}{\url{https://databricks.com/blog/2018/11/19/introducing-horovodrunner-for-distributed-deep-learning-training.html}}.
\newblock
\newblock
\shownote{Retrieved on March 16, 2022}.


\bibitem[nvi(2018)]%
        {nvidia}
 \bibinfo{year}{2018}\natexlab{}.
\newblock \bibinfo{title}{NVIDIA: Accelerating Deep Learning with Uber’s
  Horovod}.
\newblock
  \bibinfo{howpublished}{\url{https://eng.uber.com/nvidia-horovod-deep-learning/}}.
\newblock
\newblock
\shownote{Retrieved on March 16, 2022}.


\bibitem[ube(2018)]%
        {uber}
 \bibinfo{year}{2018}\natexlab{}.
\newblock \bibinfo{title}{Open Source at Uber: Meet Alex Sergeev, Horovod
  Project Lead}.
\newblock
  \bibinfo{howpublished}{\url{https://eng.uber.com/alex-sergeev-horovod/}}.
\newblock
\newblock
\shownote{Retrieved on March 16, 2022}.


\bibitem[hor(2018c)]%
        {horovod467}
 \bibinfo{year}{2018}\natexlab{c}.
\newblock \bibinfo{title}{Permission denied (publickey,password) when I run on
  muti node}.
\newblock
  \bibinfo{howpublished}{\url{https://github.com/horovod/horovod/issues/467}}.
\newblock
\newblock
\shownote{Retrieved on March 16, 2022}.


\bibitem[ope(2019)]%
        {openai}
 \bibinfo{year}{2019}\natexlab{}.
\newblock \bibinfo{title}{{AI and Compute}}.
\newblock
  \bibinfo{howpublished}{\url{https://openai.com/blog/ai-and-compute/}}.
\newblock
\newblock
\shownote{Retrieved on March 16, 2022}.


\bibitem[nvi(2019)]%
        {nvidia2}
 \bibinfo{year}{2019}\natexlab{}.
\newblock \bibinfo{title}{Distributed Deep Learning with Horovod}.
\newblock
  \bibinfo{howpublished}{\url{https://developer.download.nvidia.cn/video/gputechconf/gtc/2019/presentation/s9321-distributed-deep-learning-with-horovod.pdf}}.
\newblock
\newblock
\shownote{Retrieved on March 16, 2022}.


\bibitem[FdD(2019)]%
        {FdDL}
 \bibinfo{year}{2019}\natexlab{}.
\newblock \bibinfo{title}{Fabric for Deep Learning (FfDL)}.
\newblock \bibinfo{howpublished}{\url{https://github.com/IBM/FfDL}}.
\newblock
\newblock
\shownote{Retrieved on March 16, 2022}.


\bibitem[hor(2019)]%
        {horovod818}
 \bibinfo{year}{2019}\natexlab{}.
\newblock \bibinfo{title}{Horovod pypi release doesn't have horovod.spark
  package}.
\newblock
  \bibinfo{howpublished}{\url{https://github.com/horovod/horovod/issues/818}}.
\newblock
\newblock
\shownote{Retrieved on December 19, 2022}.


\bibitem[ncc(2019)]%
        {nccl}
 \bibinfo{year}{2019}\natexlab{}.
\newblock \bibinfo{title}{{NCCL}}.
\newblock \bibinfo{howpublished}{\url{https://developer.nvidia.com/nccl}}.
\newblock
\newblock
\shownote{Retrieved on March 16, 2022}.


\bibitem[pyt(2019a)]%
        {pytorch22834}
 \bibinfo{year}{2019}\natexlab{a}.
\newblock \bibinfo{title}{Script freezes with no output when using
  DistributedDataParallel}.
\newblock
  \bibinfo{howpublished}{\url{https://github.com/pytorch/pytorch/issues/22834}}.
\newblock
\newblock
\shownote{Retrieved on March 16, 2022}.


\bibitem[pyt(2019b)]%
        {pytorch17848}
 \bibinfo{year}{2019}\natexlab{b}.
\newblock \bibinfo{title}{torch.distributed.launch receives RuntimeError:
  ProcessGroupNCCL does not support barrier}.
\newblock
  \bibinfo{howpublished}{\url{https://github.com/pytorch/pytorch/issues/17848}}.
\newblock
\newblock
\shownote{Retrieved on March 16, 2022}.


\bibitem[so5(2019)]%
        {so57244733}
 \bibinfo{year}{2019}\natexlab{}.
\newblock \bibinfo{title}{Where does the documentation point to a list of
  values for the loss property of the compile function?}
\newblock
  \bibinfo{howpublished}{\url{https://stackoverflow.com/questions/57244733}}.
\newblock
\newblock
\shownote{Retrieved on December 21, 2022}.


\bibitem[pyt(2020)]%
        {pytorch38300}
 \bibinfo{year}{2020}\natexlab{}.
\newblock \bibinfo{title}{AssertionError: Default process group is not
  initialized}.
\newblock
  \bibinfo{howpublished}{\url{https://github.com/pytorch/pytorch/issues/38300}}.
\newblock
\newblock
\shownote{Retrieved on March 16, 2022}.


\bibitem[low(2020)]%
        {lowefficiency}
 \bibinfo{year}{2020}\natexlab{}.
\newblock \bibinfo{title}{CIFAR Scaling Efficiency}.
\newblock
  \bibinfo{howpublished}{\url{https://github.com/horovod/horovod/issues/2103}}.
\newblock
\newblock
\shownote{Retrieved on March 16, 2022}.


\bibitem[hor(2020a)]%
        {horovod1910}
 \bibinfo{year}{2020}\natexlab{a}.
\newblock \bibinfo{title}{{Installation issue with MXNet built from source - No
  such file or dictionary <dmlc/base.h>}}.
\newblock
  \bibinfo{howpublished}{\url{https://github.com/horovod/horovod/issues/1910}}.
\newblock
\newblock
\shownote{Retrieved on March 16, 2022}.


\bibitem[mpi(2020)]%
        {mpi}
 \bibinfo{year}{2020}\natexlab{}.
\newblock \bibinfo{title}{Open {MPI: Open Source High Performance Computing}}.
\newblock \bibinfo{howpublished}{\url{https://www.open-mpi.org}}.
\newblock
\newblock
\shownote{Retrieved on March 16, 2022}.


\bibitem[fra(2020)]%
        {frameworks1}
 \bibinfo{year}{2020}\natexlab{}.
\newblock \bibinfo{title}{Popular Deep Learning Frameworks: An Overview}.
\newblock
  \bibinfo{howpublished}{\url{https://analyticsindiamag.com/deep-learning-frameworks/}}.
\newblock
\newblock
\shownote{Retrieved on March 16, 2022}.


\bibitem[so6(2020a)]%
        {so60799655}
 \bibinfo{year}{2020}\natexlab{a}.
\newblock \bibinfo{title}{Pytorch DataParallel doesn't work when the model
  contain tensor operation}.
\newblock
  \bibinfo{howpublished}{\url{https://stackoverflow.com/questions/60799655}}.
\newblock
\newblock
\shownote{Retrieved on March 16, 2022}.


\bibitem[so6(2020b)]%
        {so65358676}
 \bibinfo{year}{2020}\natexlab{b}.
\newblock \bibinfo{title}{What does 'with strategy.scope():' or 'with
  tf.distribute.experimental.TPUStrategy(tpu).scope():' do to the creation of a
  NN?}
\newblock
  \bibinfo{howpublished}{\url{https://stackoverflow.com/questions/65358676}}.
\newblock
\newblock
\shownote{Retrieved on January 16, 2023}.


\bibitem[err(2020)]%
        {errormessage}
 \bibinfo{year}{2020}\natexlab{}.
\newblock \bibinfo{title}{When Build Docker Container with Ubuntu16.04 Install
  Horovod Failed with Error Code -4}.
\newblock
  \bibinfo{howpublished}{\url{https://github.com/horovod/horovod/issues/1798}}.
\newblock
\newblock
\shownote{Retrieved on March 16, 2022}.


\bibitem[hor(2020b)]%
        {horovod1798}
 \bibinfo{year}{2020}\natexlab{b}.
\newblock \bibinfo{title}{{When build docker container with ubuntu16.04 install
  horovod failed with error code -4}}.
\newblock
  \bibinfo{howpublished}{\url{https://github.com/horovod/horovod/issues/1798}}.
\newblock
\newblock
\shownote{Retrieved on March 16, 2022}.


\bibitem[ker(2021a)]%
        {keras106}
 \bibinfo{year}{2021}\natexlab{a}.
\newblock \bibinfo{title}{does it automatically use multiple gpu, if availabe?}
\newblock
  \bibinfo{howpublished}{\url{https://github.com/keras-team/keras/issues/106}}.
\newblock
\newblock
\shownote{Retrieved on Feburary 11, 2023}.


\bibitem[git(2021)]%
        {githubsearch}
 \bibinfo{year}{2021}\natexlab{}.
\newblock \bibinfo{title}{Github Search API}.
\newblock
  \bibinfo{howpublished}{\url{https://developer.github.com/v3/search/}}.
\newblock
\newblock
\shownote{Retrieved on March 16, 2022}.


\bibitem[glo(2021)]%
        {gloo}
 \bibinfo{year}{2021}\natexlab{}.
\newblock \bibinfo{title}{Gloo}.
\newblock
  \bibinfo{howpublished}{\url{https://github.com/facebookincubator/gloo}}.
\newblock
\newblock
\shownote{Retrieved on March 16, 2022}.


\bibitem[gpt(2021)]%
        {gpt3}
 \bibinfo{year}{2021}\natexlab{}.
\newblock \bibinfo{title}{{GPT-3 Powers the Next Generation of Apps}}.
\newblock \bibinfo{howpublished}{\url{https://openai.com/blog/gpt-3-apps/}}.
\newblock
\newblock
\shownote{Retrieved on March 16, 2022}.


\bibitem[hor(2021)]%
        {horovod}
 \bibinfo{year}{2021}\natexlab{}.
\newblock \bibinfo{title}{{Horovod}}.
\newblock \bibinfo{howpublished}{\url{https://github.com/horovod/horovod}}.
\newblock
\newblock
\shownote{Retrieved on March 16, 2022}.


\bibitem[rep(2021)]%
        {reporduce}
 \bibinfo{year}{2021}\natexlab{}.
\newblock \bibinfo{title}{{I meet deadlock problem when use horovod.}}
\newblock
  \bibinfo{howpublished}{\url{https://github.com/horovod/horovod/issues/2506}}.
\newblock
\newblock
\shownote{Retrieved on March 16, 2022}.


\bibitem[pyt(2021a)]%
        {pytorch54266}
 \bibinfo{year}{2021}\natexlab{a}.
\newblock \bibinfo{title}{init\_rpc: TENSOR\_PIPE backend sigaborts when CUDA
  is not available}.
\newblock
  \bibinfo{howpublished}{\url{https://github.com/pytorch/pytorch/issues/54266}}.
\newblock
\newblock
\shownote{Retrieved on December 19, 2022}.


\bibitem[ker(2021b)]%
        {keras}
 \bibinfo{year}{2021}\natexlab{b}.
\newblock \bibinfo{title}{{Keras: Deep Learning for Python}}.
\newblock \bibinfo{howpublished}{\url{https://github.com/keras-team/keras}}.
\newblock
\newblock
\shownote{Retrieved on March 16, 2022}.


\bibitem[pad(2021)]%
        {paddle}
 \bibinfo{year}{2021}\natexlab{}.
\newblock \bibinfo{title}{{PaddlePaddle}}.
\newblock \bibinfo{howpublished}{\url{https://github.com/PaddlePaddle/Paddle}}.
\newblock
\newblock
\shownote{Retrieved on March 16, 2022}.


\bibitem[pyt(2021b)]%
        {pytorch}
 \bibinfo{year}{2021}\natexlab{b}.
\newblock \bibinfo{title}{{PyTorch}}.
\newblock \bibinfo{howpublished}{\url{https://github.com/pytorch/pytorch}}.
\newblock
\newblock
\shownote{Retrieved on March 16, 2022}.


\bibitem[dro(2021)]%
        {dropout}
 \bibinfo{year}{2021}\natexlab{}.
\newblock \bibinfo{title}{Running a Basic Distributed MNIST Solver in
  TensorFlow}.
\newblock
  \bibinfo{howpublished}{\url{https://stackoverflow.com/questions/49984317/running-a-basic-distributed-mnist-solver-in-tensorflow}}.
\newblock
\newblock
\shownote{Retrieved on March 16, 2022}.


\bibitem[sta(2021)]%
        {stackexchange}
 \bibinfo{year}{2021}\natexlab{}.
\newblock \bibinfo{title}{Stack Exchange Data Dump}.
\newblock
  \bibinfo{howpublished}{\url{https://archive.org/details/stackexchange}}.
\newblock
\newblock
\shownote{Retrieved on December 6, 2021}.


\bibitem[ten(2021)]%
        {tensorflow}
 \bibinfo{year}{2021}\natexlab{}.
\newblock \bibinfo{title}{{TensorFlow}}.
\newblock
  \bibinfo{howpublished}{\url{https://github.com/tensorflow/tensorflow}}.
\newblock
\newblock
\shownote{Retrieved on March 16, 2022}.


\bibitem[fra(2021a)]%
        {frameworks4}
 \bibinfo{year}{2021}\natexlab{a}.
\newblock \bibinfo{title}{Top 10 Deep Learning Frameworks in 2021 You Can’t
  Ignore}.
\newblock
  \bibinfo{howpublished}{\url{https://www.upgrad.com/blog/top-deep-learning-frameworks/}}.
\newblock
\newblock
\shownote{Retrieved on March 16, 2022}.


\bibitem[fra(2021b)]%
        {frameworks2}
 \bibinfo{year}{2021}\natexlab{b}.
\newblock \bibinfo{title}{Top 10 Deep Learning Frameworks in 2022 You Can’t
  Ignore}.
\newblock
  \bibinfo{howpublished}{\url{https://www.upgrad.com/blog/top-deep-learning-frameworks/}}.
\newblock
\newblock
\shownote{Retrieved on March 16, 2022}.


\bibitem[fra(2021c)]%
        {frameworks3}
 \bibinfo{year}{2021}\natexlab{c}.
\newblock \bibinfo{title}{Top 5 Deep Learning Frameworks in 2021}.
\newblock
  \bibinfo{howpublished}{\url{https://makeinbusiness.com/top-5-deep-learning-frameworks/}}.
\newblock
\newblock
\shownote{Retrieved on March 16, 2022}.


\bibitem[dis(2023)]%
        {distributed_training_device}
 \bibinfo{year}{2023}\natexlab{}.
\newblock \bibinfo{title}{Distributed training of deep learning models on
  Azure}.
\newblock
  \bibinfo{howpublished}{\url{https://learn.microsoft.com/en-us/azure/architecture/reference-architectures/ai/training-deep-learning}}.
\newblock
\newblock
\shownote{Retrieved on April 18, 2023}.


\bibitem[dat(2023)]%
        {dataset}
 \bibinfo{year}{2023}\natexlab{}.
\newblock \bibinfo{title}{Supplemental Materials}.
\newblock
  \bibinfo{howpublished}{\url{https://github.com/gudiandian/TOSEM23-DistributedTraining}}.
\newblock
\newblock
\shownote{Retrieved on April 22, 2023}.


\bibitem[Abadi et~al\mbox{.}(2016)]%
        {AbadiBCCDDDGIIK16}
\bibfield{author}{\bibinfo{person}{Mart{\'{\i}}n Abadi}, \bibinfo{person}{Paul
  Barham}, \bibinfo{person}{Jianmin Chen}, \bibinfo{person}{Zhifeng Chen},
  \bibinfo{person}{Andy Davis}, \bibinfo{person}{Jeffrey Dean},
  \bibinfo{person}{Matthieu Devin}, \bibinfo{person}{Sanjay Ghemawat},
  \bibinfo{person}{Geoffrey Irving}, \bibinfo{person}{Michael Isard},
  \bibinfo{person}{Manjunath Kudlur}, \bibinfo{person}{Josh Levenberg},
  \bibinfo{person}{Rajat Monga}, \bibinfo{person}{Sherry Moore},
  \bibinfo{person}{Derek~Gordon Murray}, \bibinfo{person}{Benoit Steiner},
  \bibinfo{person}{Paul~A. Tucker}, \bibinfo{person}{Vijay Vasudevan},
  \bibinfo{person}{Pete Warden}, \bibinfo{person}{Martin Wicke},
  \bibinfo{person}{Yuan Yu}, {and} \bibinfo{person}{Xiaoqiang Zheng}.}
  \bibinfo{year}{2016}\natexlab{}.
\newblock \showarticletitle{TensorFlow: {A} System for Large-Scale Machine
  Learning}. In \bibinfo{booktitle}{\emph{Proceedings of 12th {USENIX}
  Symposium on Operating Systems Design and Implementation, {OSDI} 2016}}.
  \bibinfo{pages}{265--283}.
\newblock


\bibitem[Aghajani et~al\mbox{.}(2019)]%
        {AghajaniNVLMBL19}
\bibfield{author}{\bibinfo{person}{Emad Aghajani}, \bibinfo{person}{Csaba
  Nagy}, \bibinfo{person}{Olga~Lucero Vega{-}M{\'{a}}rquez},
  \bibinfo{person}{Mario Linares{-}V{\'{a}}squez}, \bibinfo{person}{Laura
  Moreno}, \bibinfo{person}{Gabriele Bavota}, {and} \bibinfo{person}{Michele
  Lanza}.} \bibinfo{year}{2019}\natexlab{}.
\newblock \showarticletitle{Software Documentation Issues Unveiled}. In
  \bibinfo{booktitle}{\emph{Proceedings of the 41st International Conference on
  Software Engineering, {ICSE} 2019}}. \bibinfo{pages}{1199--1210}.
\newblock


\bibitem[Asadollah(2018)]%
        {Asadollah18}
\bibfield{author}{\bibinfo{person}{Sara~Abbaspour Asadollah}.}
  \bibinfo{year}{2018}\natexlab{}.
\newblock \emph{\bibinfo{title}{Concurrency Bugs: Characterization, Debugging
  and Runtime Verification}}.
\newblock \bibinfo{thesistype}{Ph.\,D. Dissertation}.
  \bibinfo{school}{M{\"{a}}lardalen University College, V{\"{a}}ster{\aa}s,
  Eskilstuna, Sweden}.
\newblock


\bibitem[Bai et~al\mbox{.}(2021)]%
        {bai2021gradient}
\bibfield{author}{\bibinfo{person}{Youhui Bai}, \bibinfo{person}{Cheng Li},
  \bibinfo{person}{Quan Zhou}, \bibinfo{person}{Jun Yi}, \bibinfo{person}{Ping
  Gong}, \bibinfo{person}{Feng Yan}, \bibinfo{person}{Ruichuan Chen}, {and}
  \bibinfo{person}{Yinlong Xu}.} \bibinfo{year}{2021}\natexlab{}.
\newblock \showarticletitle{Gradient compression supercharged high-performance
  data parallel dnn training}. In \bibinfo{booktitle}{\emph{Proceedings of the
  ACM SIGOPS 28th Symposium on Operating Systems Principles, SIGOPS 2021}}.
  \bibinfo{pages}{359--375}.
\newblock


\bibitem[Ben{-}Nun and Hoefler(2019)]%
        {Ben-NunH19}
\bibfield{author}{\bibinfo{person}{Tal Ben{-}Nun} {and}
  \bibinfo{person}{Torsten Hoefler}.} \bibinfo{year}{2019}\natexlab{}.
\newblock \showarticletitle{Demystifying Parallel and Distributed Deep
  Learning: An In-depth Concurrency Analysis}.
\newblock \bibinfo{journal}{\emph{{ACM} Comput. Surv.}} \bibinfo{volume}{52},
  \bibinfo{number}{4} (\bibinfo{year}{2019}), \bibinfo{pages}{65:1--65:43}.
\newblock


\bibitem[Berard et~al\mbox{.}(2016)]%
        {BerardPSB16}
\bibfield{author}{\bibinfo{person}{Alexandre Berard}, \bibinfo{person}{Olivier
  Pietquin}, \bibinfo{person}{Christophe Servan}, {and}
  \bibinfo{person}{Laurent Besacier}.} \bibinfo{year}{2016}\natexlab{}.
\newblock \showarticletitle{Listen and Translate: {A} Proof of Concept for
  End-to-End Speech-to-Text Translation}.
\newblock \bibinfo{journal}{\emph{CoRR}}  \bibinfo{volume}{abs/1612.01744}
  (\bibinfo{year}{2016}).
\newblock


\bibitem[Brennan and Prediger(1981)]%
        {brennan1981coefficient}
\bibfield{author}{\bibinfo{person}{Robert~L Brennan} {and}
  \bibinfo{person}{Dale~J Prediger}.} \bibinfo{year}{1981}\natexlab{}.
\newblock \showarticletitle{Coefficient Kappa: Some Uses, Misuses, and
  Alternatives}.
\newblock \bibinfo{journal}{\emph{Educational and psychological measurement}}
  \bibinfo{volume}{41}, \bibinfo{number}{3} (\bibinfo{year}{1981}),
  \bibinfo{pages}{687--699}.
\newblock


\bibitem[Chen et~al\mbox{.}(2015)]%
        {ChenSKX15}
\bibfield{author}{\bibinfo{person}{Chenyi Chen}, \bibinfo{person}{Ari Seff},
  \bibinfo{person}{Alain~L. Kornhauser}, {and} \bibinfo{person}{Jianxiong
  Xiao}.} \bibinfo{year}{2015}\natexlab{}.
\newblock \showarticletitle{DeepDriving: Learning Affordance for Direct
  Perception in Autonomous Driving}. In \bibinfo{booktitle}{\emph{Proceedings
  of 2015 {IEEE} International Conference on Computer Vision, {ICCV} 2015}}.
  \bibinfo{pages}{2722--2730}.
\newblock


\bibitem[Chen et~al\mbox{.}(2020)]%
        {ChenCLW0L20}
\bibfield{author}{\bibinfo{person}{Zhenpeng Chen}, \bibinfo{person}{Yanbin
  Cao}, \bibinfo{person}{Yuanqiang Liu}, \bibinfo{person}{Haoyu Wang},
  \bibinfo{person}{Tao Xie}, {and} \bibinfo{person}{Xuanzhe Liu}.}
  \bibinfo{year}{2020}\natexlab{}.
\newblock \showarticletitle{A Comprehensive Study on Challenges in Deploying
  Deep Learning Based Software}. In \bibinfo{booktitle}{\emph{Proceedings of
  28th {ACM} Joint European Software Engineering Conference and Symposium on
  the Foundations of Software Engineering, {ESEC/SIGSOFT} {FSE} 2020}}.
  \bibinfo{pages}{750--762}.
\newblock


\bibitem[Chen et~al\mbox{.}(2021)]%
        {abs-2101-04930}
\bibfield{author}{\bibinfo{person}{Zhenpeng Chen}, \bibinfo{person}{Huihan
  Yao}, \bibinfo{person}{Yiling Lou}, \bibinfo{person}{Yanbin Cao},
  \bibinfo{person}{Yuanqiang Liu}, \bibinfo{person}{Haoyu Wang}, {and}
  \bibinfo{person}{Xuanzhe Liu}.} \bibinfo{year}{2021}\natexlab{}.
\newblock \showarticletitle{An Empirical Study on Deployment Faults of Deep
  Learning Based Mobile Applications}. In \bibinfo{booktitle}{\emph{Proceedings
  of the 43rd International Conference on Software Engineering, ICSE 2021}}.
  \bibinfo{pages}{674--685}.
\newblock


\bibitem[Cohen(1960)]%
        {cohen1960coefficient}
\bibfield{author}{\bibinfo{person}{Jacob Cohen}.}
  \bibinfo{year}{1960}\natexlab{}.
\newblock \showarticletitle{A Coefficient of Agreement for Nominal Scales}.
\newblock \bibinfo{journal}{\emph{Educational and psychological measurement}}
  \bibinfo{volume}{20}, \bibinfo{number}{1} (\bibinfo{year}{1960}),
  \bibinfo{pages}{37--46}.
\newblock


\bibitem[Dean et~al\mbox{.}(2012)]%
        {DeanCMCDLMRSTYN12}
\bibfield{author}{\bibinfo{person}{Jeffrey Dean}, \bibinfo{person}{Greg
  Corrado}, \bibinfo{person}{Rajat Monga}, \bibinfo{person}{Kai Chen},
  \bibinfo{person}{Matthieu Devin}, \bibinfo{person}{Quoc~V. Le},
  \bibinfo{person}{Mark~Z. Mao}, \bibinfo{person}{Marc'Aurelio Ranzato},
  \bibinfo{person}{Andrew~W. Senior}, \bibinfo{person}{Paul~A. Tucker},
  \bibinfo{person}{Ke Yang}, {and} \bibinfo{person}{Andrew~Y. Ng}.}
  \bibinfo{year}{2012}\natexlab{}.
\newblock \showarticletitle{Large Scale Distributed Deep Networks}. In
  \bibinfo{booktitle}{\emph{Proceedings of 26th Annual Conference on Neural
  Information Processing Systems, NeurIPS 2012}}. \bibinfo{pages}{1232--1240}.
\newblock


\bibitem[Devlin et~al\mbox{.}(2019)]%
        {DevlinCLT19}
\bibfield{author}{\bibinfo{person}{Jacob Devlin}, \bibinfo{person}{Ming{-}Wei
  Chang}, \bibinfo{person}{Kenton Lee}, {and} \bibinfo{person}{Kristina
  Toutanova}.} \bibinfo{year}{2019}\natexlab{}.
\newblock \showarticletitle{{BERT:} Pre-training of Deep Bidirectional
  Transformers for Language Understanding}. In
  \bibinfo{booktitle}{\emph{Proceedings of the 2019 Conference of the North
  American Chapter of the Association for Computational Linguistics: Human
  Language Technologies, {NAACL-HLT} 2019}}. \bibinfo{pages}{4171--4186}.
\newblock


\bibitem[Ferreira and Zwinderman(2006)]%
        {ferreira2006benjamini}
\bibfield{author}{\bibinfo{person}{JA Ferreira} {and} \bibinfo{person}{AH
  Zwinderman}.} \bibinfo{year}{2006}\natexlab{}.
\newblock \showarticletitle{On The Benjamini--Hochberg Method}.
\newblock \bibinfo{journal}{\emph{The Annals of Statistics}}
  \bibinfo{volume}{34}, \bibinfo{number}{4} (\bibinfo{year}{2006}),
  \bibinfo{pages}{1827--1849}.
\newblock


\bibitem[Fogel et~al\mbox{.}(2015)]%
        {FogelFPWGMM15}
\bibfield{author}{\bibinfo{person}{Ari Fogel}, \bibinfo{person}{Stanley Fung},
  \bibinfo{person}{Luis Pedrosa}, \bibinfo{person}{Meg Walraed{-}Sullivan},
  \bibinfo{person}{Ramesh Govindan}, \bibinfo{person}{Ratul Mahajan}, {and}
  \bibinfo{person}{Todd~D. Millstein}.} \bibinfo{year}{2015}\natexlab{}.
\newblock \showarticletitle{A General Approach to Network Configuration
  Analysis}. In \bibinfo{booktitle}{\emph{Proceedings of 12th {USENIX}
  Symposium on Networked Systems Design and Implementation, {NSDI} 2015}}.
  \bibinfo{pages}{469--483}.
\newblock


\bibitem[Franco et~al\mbox{.}(2017)]%
        {FrancoGR17}
\bibfield{author}{\bibinfo{person}{Anthony~Di Franco}, \bibinfo{person}{Hui
  Guo}, {and} \bibinfo{person}{Cindy Rubio{-}Gonz{\'{a}}lez}.}
  \bibinfo{year}{2017}\natexlab{}.
\newblock \showarticletitle{A Comprehensive Study of Real-World Numerical Bug
  Characteristics}. In \bibinfo{booktitle}{\emph{Proceedings of the 32nd
  {IEEE/ACM} International Conference on Automated Software Engineering, {ASE}
  2017}}. \bibinfo{pages}{509--519}.
\newblock


\bibitem[Gao et~al\mbox{.}(2018)]%
        {GaoDQGW0HZW18}
\bibfield{author}{\bibinfo{person}{Yu Gao}, \bibinfo{person}{Wensheng Dou},
  \bibinfo{person}{Feng Qin}, \bibinfo{person}{Chushu Gao},
  \bibinfo{person}{Dong Wang}, \bibinfo{person}{Jun Wei},
  \bibinfo{person}{Ruirui Huang}, \bibinfo{person}{Li Zhou}, {and}
  \bibinfo{person}{Yongming Wu}.} \bibinfo{year}{2018}\natexlab{}.
\newblock \showarticletitle{An Empirical Study on Crash Recovery Bugs in
  Large-Scale Distributed Systems}. In \bibinfo{booktitle}{\emph{Proceedings of
  the 2018 {ACM} Joint Meeting on European Software Engineering Conference and
  Symposium on the Foundations of Software Engineering, {ESEC/SIGSOFT} {FSE}
  2018}}. \bibinfo{pages}{539--550}.
\newblock


\bibitem[Greenwood and Nikulin(1996)]%
        {greenwood1996guide}
\bibfield{author}{\bibinfo{person}{Priscilla~E Greenwood} {and}
  \bibinfo{person}{Michael~S Nikulin}.} \bibinfo{year}{1996}\natexlab{}.
\newblock \bibinfo{booktitle}{\emph{A Guide to Chi-Aquared Testing}}.
  Vol.~\bibinfo{volume}{280}.
\newblock


\bibitem[Gu et~al\mbox{.}(2023)]%
        {Gu:AsPLOS23}
\bibfield{author}{\bibinfo{person}{Diandian Gu}, \bibinfo{person}{Yihao Zhao},
  \bibinfo{person}{Yinmin Zhong}, \bibinfo{person}{Yifan Xiong},
  \bibinfo{person}{Zhenhua Han}, \bibinfo{person}{Peng Cheng},
  \bibinfo{person}{Fan Yang}, \bibinfo{person}{Gang Huang},
  \bibinfo{person}{Xin Jin}, {and} \bibinfo{person}{Xuanzhe Liu}.}
  \bibinfo{year}{2023}\natexlab{}.
\newblock \showarticletitle{ElasticFlow: An Elastic Serverless Training
  Platform for Distributed Deep Learning}. In
  \bibinfo{booktitle}{\emph{Proceedings of the 28th {ACM} International
  Conference on Architectural Support for Programming Languages and Operating
  Systems, Volume 2, {ASPLOS} 2023, Vancouver, BC, Canada, March 25-29, 2023}}.
  \bibinfo{publisher}{{ACM}}, \bibinfo{pages}{266--280}.
\newblock


\bibitem[Hazelwood et~al\mbox{.}(2018)]%
        {HazelwoodBBCDDF18}
\bibfield{author}{\bibinfo{person}{Kim~M. Hazelwood}, \bibinfo{person}{Sarah
  Bird}, \bibinfo{person}{David~M. Brooks}, \bibinfo{person}{Soumith Chintala},
  \bibinfo{person}{Utku Diril}, \bibinfo{person}{Dmytro Dzhulgakov},
  \bibinfo{person}{Mohamed Fawzy}, \bibinfo{person}{Bill Jia},
  \bibinfo{person}{Yangqing Jia}, \bibinfo{person}{Aditya Kalro},
  \bibinfo{person}{James Law}, \bibinfo{person}{Kevin Lee},
  \bibinfo{person}{Jason Lu}, \bibinfo{person}{Pieter Noordhuis},
  \bibinfo{person}{Misha Smelyanskiy}, \bibinfo{person}{Liang Xiong}, {and}
  \bibinfo{person}{Xiaodong Wang}.} \bibinfo{year}{2018}\natexlab{}.
\newblock \showarticletitle{Applied Machine Learning at Facebook: {A}
  Datacenter Infrastructure Perspective}. In
  \bibinfo{booktitle}{\emph{Proceedings of {IEEE} International Symposium on
  High Performance Computer Architecture, {HPCA} 2018}}.
  \bibinfo{pages}{620--629}.
\newblock


\bibitem[He et~al\mbox{.}(2016)]%
        {HeZRS16}
\bibfield{author}{\bibinfo{person}{Kaiming He}, \bibinfo{person}{Xiangyu
  Zhang}, \bibinfo{person}{Shaoqing Ren}, {and} \bibinfo{person}{Jian Sun}.}
  \bibinfo{year}{2016}\natexlab{}.
\newblock \showarticletitle{Deep Residual Learning for Image Recognition}. In
  \bibinfo{booktitle}{\emph{Proceedings of 2016 {IEEE} Conference on Computer
  Vision and Pattern Recognition, {CVPR} 2016}}. \bibinfo{pages}{770--778}.
\newblock


\bibitem[Huang et~al\mbox{.}(2019)]%
        {HuangCBFCCLNLWC19}
\bibfield{author}{\bibinfo{person}{Yanping Huang}, \bibinfo{person}{Youlong
  Cheng}, \bibinfo{person}{Ankur Bapna}, \bibinfo{person}{Orhan Firat},
  \bibinfo{person}{Dehao Chen}, \bibinfo{person}{Mia~Xu Chen},
  \bibinfo{person}{HyoukJoong Lee}, \bibinfo{person}{Jiquan Ngiam},
  \bibinfo{person}{Quoc~V. Le}, \bibinfo{person}{Yonghui Wu}, {and}
  \bibinfo{person}{Zhifeng Chen}.} \bibinfo{year}{2019}\natexlab{}.
\newblock \showarticletitle{GPipe: Efficient Training of Giant Neural Networks
  using Pipeline Parallelism}. In \bibinfo{booktitle}{\emph{Proceedings of 32nd
  Annual Conference on Neural Information Processing Systems, NeurIPS 2019}}.
  \bibinfo{pages}{103--112}.
\newblock


\bibitem[Humbatova et~al\mbox{.}(2020)]%
        {HumbatovaJBR0T20}
\bibfield{author}{\bibinfo{person}{Nargiz Humbatova}, \bibinfo{person}{Gunel
  Jahangirova}, \bibinfo{person}{Gabriele Bavota}, \bibinfo{person}{Vincenzo
  Riccio}, \bibinfo{person}{Andrea Stocco}, {and} \bibinfo{person}{Paolo
  Tonella}.} \bibinfo{year}{2020}\natexlab{}.
\newblock \showarticletitle{Taxonomy of Real Faults in Deep Learning Systems}.
  In \bibinfo{booktitle}{\emph{Proceedings of 42nd International Conference on
  Software Engineering, {ICSE} 2020}}. \bibinfo{pages}{1110--1121}.
\newblock


\bibitem[Islam et~al\mbox{.}(2019)]%
        {IslamNPR19}
\bibfield{author}{\bibinfo{person}{Md~Johirul Islam}, \bibinfo{person}{Giang
  Nguyen}, \bibinfo{person}{Rangeet Pan}, {and} \bibinfo{person}{Hridesh
  Rajan}.} \bibinfo{year}{2019}\natexlab{}.
\newblock \showarticletitle{A Comprehensive Study on Deep Learning Bug
  Characteristics}. In \bibinfo{booktitle}{\emph{Proceedings of 27th {ACM}
  Joint Meeting on European Software Engineering Conference and Symposium on
  the Foundations of Software Engineering, {ESEC/SIGSOFT} {FSE} 2019}}.
  \bibinfo{pages}{510--520}.
\newblock


\bibitem[Islam et~al\mbox{.}(2020)]%
        {IslamPNR20}
\bibfield{author}{\bibinfo{person}{Md~Johirul Islam}, \bibinfo{person}{Rangeet
  Pan}, \bibinfo{person}{Giang Nguyen}, {and} \bibinfo{person}{Hridesh Rajan}.}
  \bibinfo{year}{2020}\natexlab{}.
\newblock \showarticletitle{Repairing Deep Neural Networks: Fix Patterns and
  Challenges}. In \bibinfo{booktitle}{\emph{Proceedings of 42nd International
  Conference on Software Engineering, {ICSE} 2020}}.
  \bibinfo{pages}{1135--1146}.
\newblock


\bibitem[Jeon et~al\mbox{.}(2019)]%
        {JeonVPQXY19}
\bibfield{author}{\bibinfo{person}{Myeongjae Jeon}, \bibinfo{person}{Shivaram
  Venkataraman}, \bibinfo{person}{Amar Phanishayee}, \bibinfo{person}{Junjie
  Qian}, \bibinfo{person}{Wencong Xiao}, {and} \bibinfo{person}{Fan Yang}.}
  \bibinfo{year}{2019}\natexlab{}.
\newblock \showarticletitle{Analysis of Large-Scale Multi-Tenant {GPU} Clusters
  for {DNN} Training Workloads}. In \bibinfo{booktitle}{\emph{Proceedings of
  2019 {USENIX} Annual Technical Conference, {USENIX} {ATC} 2019}}.
  \bibinfo{pages}{947--960}.
\newblock


\bibitem[Jia et~al\mbox{.}(2019)]%
        {JiaZA19}
\bibfield{author}{\bibinfo{person}{Zhihao Jia}, \bibinfo{person}{Matei
  Zaharia}, {and} \bibinfo{person}{Alex Aiken}.}
  \bibinfo{year}{2019}\natexlab{}.
\newblock \showarticletitle{Beyond Data and Model Parallelism for Deep Neural
  Networks}. In \bibinfo{booktitle}{\emph{Proceedings of Machine Learning and
  Systems 2019, MLSys 2019}}.
\newblock


\bibitem[Jiang et~al\mbox{.}(2020)]%
        {JiangZLYCG20}
\bibfield{author}{\bibinfo{person}{Yimin Jiang}, \bibinfo{person}{Yibo Zhu},
  \bibinfo{person}{Chang Lan}, \bibinfo{person}{Bairen Yi},
  \bibinfo{person}{Yong Cui}, {and} \bibinfo{person}{Chuanxiong Guo}.}
  \bibinfo{year}{2020}\natexlab{}.
\newblock \showarticletitle{A Unified Architecture for Accelerating Distributed
  {DNN} Training in Heterogeneous {GPU/CPU} Clusters}. In
  \bibinfo{booktitle}{\emph{Proceedings of 14th {USENIX} Symposium on Operating
  Systems Design and Implementation, {OSDI} 2020}}. \bibinfo{pages}{463--479}.
\newblock


\bibitem[Jonas et~al\mbox{.}(2019)]%
        {jonas2019cloud}
\bibfield{author}{\bibinfo{person}{Eric Jonas}, \bibinfo{person}{Johann
  Schleier-Smith}, \bibinfo{person}{Vikram Sreekanti},
  \bibinfo{person}{Chia-Che Tsai}, \bibinfo{person}{Anurag Khandelwal},
  \bibinfo{person}{Qifan Pu}, \bibinfo{person}{Vaishaal Shankar},
  \bibinfo{person}{Joao Carreira}, \bibinfo{person}{Karl Krauth},
  \bibinfo{person}{Neeraja Yadwadkar}, {et~al\mbox{.}}}
  \bibinfo{year}{2019}\natexlab{}.
\newblock \showarticletitle{Cloud Programming Simplified: A Berkeley View on
  Serverless Computing}.
\newblock \bibinfo{journal}{\emph{arXiv preprint arXiv:1902.03383}}
  (\bibinfo{year}{2019}).
\newblock


\bibitem[Krizhevsky et~al\mbox{.}(2012)]%
        {KrizhevskySH12}
\bibfield{author}{\bibinfo{person}{Alex Krizhevsky}, \bibinfo{person}{Ilya
  Sutskever}, {and} \bibinfo{person}{Geoffrey~E. Hinton}.}
  \bibinfo{year}{2012}\natexlab{}.
\newblock \showarticletitle{ImageNet Classification with Deep Convolutional
  Neural Networks}. In \bibinfo{booktitle}{\emph{Proceedings of 26th Annual
  Conference on Neural Information Processing Systems, NeurIPS 2012}}.
  \bibinfo{pages}{1106--1114}.
\newblock


\bibitem[Landis and Koch(1977)]%
        {landis1977measurement}
\bibfield{author}{\bibinfo{person}{J~Richard Landis} {and}
  \bibinfo{person}{Gary~G Koch}.} \bibinfo{year}{1977}\natexlab{}.
\newblock \showarticletitle{The Measurement of Observer Agreement for
  Categorical Data}.
\newblock \bibinfo{journal}{\emph{biometrics}} (\bibinfo{year}{1977}),
  \bibinfo{pages}{159--174}.
\newblock


\bibitem[Li et~al\mbox{.}(2014)]%
        {LiAPSAJLSS14}
\bibfield{author}{\bibinfo{person}{Mu Li}, \bibinfo{person}{David~G. Andersen},
  \bibinfo{person}{Jun~Woo Park}, \bibinfo{person}{Alexander~J. Smola},
  \bibinfo{person}{Amr Ahmed}, \bibinfo{person}{Vanja Josifovski},
  \bibinfo{person}{James Long}, \bibinfo{person}{Eugene~J. Shekita}, {and}
  \bibinfo{person}{Bor{-}Yiing Su}.} \bibinfo{year}{2014}\natexlab{}.
\newblock \showarticletitle{Scaling Distributed Machine Learning with the
  Parameter Server}. In \bibinfo{booktitle}{\emph{Proceedings of 11th {USENIX}
  Symposium on Operating Systems Design and Implementation, {OSDI} 2014}}.
  \bibinfo{pages}{583--598}.
\newblock


\bibitem[Li et~al\mbox{.}(2013)]%
        {LiZLXLLX13}
\bibfield{author}{\bibinfo{person}{Sihan Li}, \bibinfo{person}{Hucheng Zhou},
  \bibinfo{person}{Haoxiang Lin}, \bibinfo{person}{Tian Xiao},
  \bibinfo{person}{Haibo Lin}, \bibinfo{person}{Wei Lin}, {and}
  \bibinfo{person}{Tao Xie}.} \bibinfo{year}{2013}\natexlab{}.
\newblock \showarticletitle{A Characteristic Study on Failures of Production
  Distributed Data-Parallel Programs}. In \bibinfo{booktitle}{\emph{Proceedings
  of 35th International Conference on Software Engineering, {ICSE} 2013}}.
  \bibinfo{pages}{963--972}.
\newblock


\bibitem[Liu et~al\mbox{.}(2014)]%
        {Liu0ZMB14}
\bibfield{author}{\bibinfo{person}{Xuanzhe Liu}, \bibinfo{person}{Gang Huang},
  \bibinfo{person}{Qi Zhao}, \bibinfo{person}{Hong Mei}, {and}
  \bibinfo{person}{M.~Brian Blake}.} \bibinfo{year}{2014}\natexlab{}.
\newblock \showarticletitle{iMashup: a Mashup-Based Framework for Service
  Composition}.
\newblock \bibinfo{journal}{\emph{Sci. China Inf. Sci.}} \bibinfo{volume}{57},
  \bibinfo{number}{1} (\bibinfo{year}{2014}), \bibinfo{pages}{1--20}.
\newblock


\bibitem[Lou et~al\mbox{.}(2020)]%
        {LouCCHZ20}
\bibfield{author}{\bibinfo{person}{Yiling Lou}, \bibinfo{person}{Zhenpeng
  Chen}, \bibinfo{person}{Yanbin Cao}, \bibinfo{person}{Dan Hao}, {and}
  \bibinfo{person}{Lu Zhang}.} \bibinfo{year}{2020}\natexlab{}.
\newblock \showarticletitle{Understanding Build Issue Resolution in Practice:
  Symptoms and Fix Patterns}. In \bibinfo{booktitle}{\emph{Proceedings of 28th
  {ACM} Joint European Software Engineering Conference and Symposium on the
  Foundations of Software Engineering, {ESEC/SIGSOFT} {FSE} 2020}}.
  \bibinfo{pages}{617--628}.
\newblock


\bibitem[Mayer and Jacobsen(2020)]%
        {MayerJ20}
\bibfield{author}{\bibinfo{person}{Ruben Mayer} {and}
  \bibinfo{person}{Hans{-}Arno Jacobsen}.} \bibinfo{year}{2020}\natexlab{}.
\newblock \showarticletitle{Scalable Deep Learning on Distributed
  Infrastructures: Challenges, Techniques, and Tools}.
\newblock \bibinfo{journal}{\emph{{ACM} Comput. Surv.}} \bibinfo{volume}{53},
  \bibinfo{number}{1} (\bibinfo{year}{2020}), \bibinfo{pages}{3:1--3:37}.
\newblock


\bibitem[Narayanan et~al\mbox{.}(2019)]%
        {narayanan2019pipedream}
\bibfield{author}{\bibinfo{person}{Deepak Narayanan}, \bibinfo{person}{Aaron
  Harlap}, \bibinfo{person}{Amar Phanishayee}, \bibinfo{person}{Vivek
  Seshadri}, \bibinfo{person}{Nikhil~R Devanur}, \bibinfo{person}{Gregory~R
  Ganger}, \bibinfo{person}{Phillip~B Gibbons}, {and} \bibinfo{person}{Matei
  Zaharia}.} \bibinfo{year}{2019}\natexlab{}.
\newblock \showarticletitle{PipeDream: Generalized Pipeline Parallelism for DNN
  Training}. In \bibinfo{booktitle}{\emph{Proceedings of the 27th ACM Symposium
  on Operating Systems Principles, {SOSP} 2019}}. \bibinfo{pages}{1--15}.
\newblock


\bibitem[Nickolls et~al\mbox{.}(2008)]%
        {cuda}
\bibfield{author}{\bibinfo{person}{John Nickolls}, \bibinfo{person}{Ian Buck},
  \bibinfo{person}{Michael Garland}, {and} \bibinfo{person}{Kevin Skadron}.}
  \bibinfo{year}{2008}\natexlab{}.
\newblock \showarticletitle{Scalable Parallel Programming with {CUDA}}.
\newblock \bibinfo{journal}{\emph{{ACM} Queue}} \bibinfo{volume}{6},
  \bibinfo{number}{2} (\bibinfo{year}{2008}), \bibinfo{pages}{40--53}.
\newblock


\bibitem[Paszke et~al\mbox{.}(2019)]%
        {PaszkeGMLBCKLGA19}
\bibfield{author}{\bibinfo{person}{Adam Paszke}, \bibinfo{person}{Sam Gross},
  \bibinfo{person}{Francisco Massa}, \bibinfo{person}{Adam Lerer},
  \bibinfo{person}{James Bradbury}, \bibinfo{person}{Gregory Chanan},
  \bibinfo{person}{Trevor Killeen}, \bibinfo{person}{Zeming Lin},
  \bibinfo{person}{Natalia Gimelshein}, \bibinfo{person}{Luca Antiga},
  \bibinfo{person}{Alban Desmaison}, \bibinfo{person}{Andreas K{\"{o}}pf},
  \bibinfo{person}{Edward Yang}, \bibinfo{person}{Zachary DeVito},
  \bibinfo{person}{Martin Raison}, \bibinfo{person}{Alykhan Tejani},
  \bibinfo{person}{Sasank Chilamkurthy}, \bibinfo{person}{Benoit Steiner},
  \bibinfo{person}{Lu Fang}, \bibinfo{person}{Junjie Bai}, {and}
  \bibinfo{person}{Soumith Chintala}.} \bibinfo{year}{2019}\natexlab{}.
\newblock \showarticletitle{PyTorch: An Imperative Style, High-Performance Deep
  Learning Library}. In \bibinfo{booktitle}{\emph{Proceedings of 32nd Annual
  Conference on Neural Information Processing Systems, NeurIPS 2019}}.
  \bibinfo{pages}{8024--8035}.
\newblock


\bibitem[Peng et~al\mbox{.}(2019)]%
        {PengZCBYLWG19}
\bibfield{author}{\bibinfo{person}{Yanghua Peng}, \bibinfo{person}{Yibo Zhu},
  \bibinfo{person}{Yangrui Chen}, \bibinfo{person}{Yixin Bao},
  \bibinfo{person}{Bairen Yi}, \bibinfo{person}{Chang Lan},
  \bibinfo{person}{Chuan Wu}, {and} \bibinfo{person}{Chuanxiong Guo}.}
  \bibinfo{year}{2019}\natexlab{}.
\newblock \showarticletitle{A Generic Communication Scheduler for Distributed
  {DNN} Training Acceleration}. In \bibinfo{booktitle}{\emph{Proceedings of the
  27th {ACM} Symposium on Operating Systems Principles, {SOSP} 2019}}.
  \bibinfo{pages}{16--29}.
\newblock


\bibitem[Randolph(2005)]%
        {randolph2005free}
\bibfield{author}{\bibinfo{person}{Justus~J Randolph}.}
  \bibinfo{year}{2005}\natexlab{}.
\newblock \showarticletitle{Free-Marginal Multirater Kappa (multirater
  $\kappa$free): An Alternative to Fleiss' Fixed-Marginal Multirater Kappa.}
\newblock \bibinfo{journal}{\emph{Online submission}} (\bibinfo{year}{2005}).
\newblock


\bibitem[Seaman(1999)]%
        {Seaman99}
\bibfield{author}{\bibinfo{person}{Carolyn~B. Seaman}.}
  \bibinfo{year}{1999}\natexlab{}.
\newblock \showarticletitle{Qualitative Methods in Empirical Studies of
  Software Engineering}.
\newblock \bibinfo{journal}{\emph{{IEEE} Trans. Software Eng.}}
  \bibinfo{volume}{25}, \bibinfo{number}{4} (\bibinfo{year}{1999}),
  \bibinfo{pages}{557--572}.
\newblock


\bibitem[Sergeev and Balso(2018)]%
        {abs-1802-05799}
\bibfield{author}{\bibinfo{person}{Alexander Sergeev} {and}
  \bibinfo{person}{Mike~Del Balso}.} \bibinfo{year}{2018}\natexlab{}.
\newblock \showarticletitle{Horovod: Fast and Easy Distributed Deep Learning in
  TensorFlow}.
\newblock \bibinfo{journal}{\emph{CoRR}}  \bibinfo{volume}{abs/1802.05799}
  (\bibinfo{year}{2018}).
\newblock


\bibitem[Verbraeken et~al\mbox{.}(2020)]%
        {VerbraekenWKKVR20}
\bibfield{author}{\bibinfo{person}{Joost Verbraeken}, \bibinfo{person}{Matthijs
  Wolting}, \bibinfo{person}{Jonathan Katzy}, \bibinfo{person}{Jeroen
  Kloppenburg}, \bibinfo{person}{Tim Verbelen}, {and} \bibinfo{person}{Jan~S.
  Rellermeyer}.} \bibinfo{year}{2020}\natexlab{}.
\newblock \showarticletitle{A Survey on Distributed Machine Learning}.
\newblock \bibinfo{journal}{\emph{{ACM} Comput. Surv.}} \bibinfo{volume}{53},
  \bibinfo{number}{2} (\bibinfo{year}{2020}), \bibinfo{pages}{30:1--30:33}.
\newblock


\bibitem[Wang et~al\mbox{.}(2019)]%
        {WangLNMMTS19}
\bibfield{author}{\bibinfo{person}{Stephanie Wang}, \bibinfo{person}{John
  Liagouris}, \bibinfo{person}{Robert Nishihara}, \bibinfo{person}{Philipp
  Moritz}, \bibinfo{person}{Ujval Misra}, \bibinfo{person}{Alexey Tumanov},
  {and} \bibinfo{person}{Ion Stoica}.} \bibinfo{year}{2019}\natexlab{}.
\newblock \showarticletitle{Lineage Stash: Fault Tolerance off the Critical
  Path}. In \bibinfo{booktitle}{\emph{Proceedings of the 27th {ACM} Symposium
  on Operating Systems Principles, {SOSP} 2019}}. \bibinfo{pages}{338--352}.
\newblock


\bibitem[Wen et~al\mbox{.}(2021)]%
        {sigsoftWenCLL00JL21}
\bibfield{author}{\bibinfo{person}{Jinfeng Wen}, \bibinfo{person}{Zhenpeng
  Chen}, \bibinfo{person}{Yi Liu}, \bibinfo{person}{Yiling Lou},
  \bibinfo{person}{Yun Ma}, \bibinfo{person}{Gang Huang}, \bibinfo{person}{Xin
  Jin}, {and} \bibinfo{person}{Xuanzhe Liu}.} \bibinfo{year}{2021}\natexlab{}.
\newblock \showarticletitle{An Empirical Study on Challenges of Application
  Development in Serverless Computing}. In
  \bibinfo{booktitle}{\emph{Proceedings of the 29th {ACM} Joint European
  Software Engineering Conference and Symposium on the Foundations of Software
  Engineering, {ESEC/FSE} 2021}}. \bibinfo{pages}{416--428}.
\newblock


\bibitem[Weng et~al\mbox{.}(2022)]%
        {MLaaS}
\bibfield{author}{\bibinfo{person}{Qizhen Weng}, \bibinfo{person}{Wencong
  Xiao}, \bibinfo{person}{Yinghao Yu}, \bibinfo{person}{Wei Wang},
  \bibinfo{person}{Cheng Wang}, \bibinfo{person}{Jian He},
  \bibinfo{person}{Yong Li}, \bibinfo{person}{Liping Zhang},
  \bibinfo{person}{Wei Lin}, \bibinfo{person}{}, {and} \bibinfo{person}{Yu
  Ding}.} \bibinfo{year}{2022}\natexlab{}.
\newblock \showarticletitle{{MLaaS} in the Wild: {Workload} Analysis and
  Scheduling in Large-Scale Heterogeneous GPU Clusters}. In
  \bibinfo{booktitle}{\emph{Proceedings of 19th USENIX Symposium on Networked
  Systems Design and Implementation, {NSDI} 2022}}.
\newblock


\bibitem[Zhang et~al\mbox{.}(2020b)]%
        {ZhangXZLLY20}
\bibfield{author}{\bibinfo{person}{Ru Zhang}, \bibinfo{person}{Wencong Xiao},
  \bibinfo{person}{Hongyu Zhang}, \bibinfo{person}{Yu Liu},
  \bibinfo{person}{Haoxiang Lin}, {and} \bibinfo{person}{Mao Yang}.}
  \bibinfo{year}{2020}\natexlab{b}.
\newblock \showarticletitle{An Empirical Study on Program Failures of Deep
  Learning Jobs}. In \bibinfo{booktitle}{\emph{Proceedings of 42nd
  International Conference on Software Engineering, {ICSE} 2020}}.
  \bibinfo{pages}{1159--1170}.
\newblock


\bibitem[Zhang et~al\mbox{.}(2018)]%
        {ZhangCCXZ18}
\bibfield{author}{\bibinfo{person}{Yuhao Zhang}, \bibinfo{person}{Yifan Chen},
  \bibinfo{person}{Shing{-}Chi Cheung}, \bibinfo{person}{Yingfei Xiong}, {and}
  \bibinfo{person}{Lu Zhang}.} \bibinfo{year}{2018}\natexlab{}.
\newblock \showarticletitle{An Empirical Study on TensorFlow Program Bugs}. In
  \bibinfo{booktitle}{\emph{Proceedings of the 27th {ACM} {SIGSOFT}
  International Symposium on Software Testing and Analysis, {ISSTA} 2018}}.
  \bibinfo{pages}{129--140}.
\newblock


\bibitem[Zhang et~al\mbox{.}(2020a)]%
        {ZhangCLWAJ20}
\bibfield{author}{\bibinfo{person}{Zhen Zhang}, \bibinfo{person}{Chaokun
  Chang}, \bibinfo{person}{Haibin Lin}, \bibinfo{person}{Yida Wang},
  \bibinfo{person}{Raman Arora}, {and} \bibinfo{person}{Xin Jin}.}
  \bibinfo{year}{2020}\natexlab{a}.
\newblock \showarticletitle{Is Network the Bottleneck of Distributed
  Training?}. In \bibinfo{booktitle}{\emph{Proceedings of the 2020 Workshop on
  Network Meets {AI} {\&} ML, {NetAI@SIGCOMM} 2020}}. \bibinfo{pages}{8--13}.
\newblock


\bibitem[Zheng et~al\mbox{.}(2022)]%
        {Zheng:SIGCOMM22}
\bibfield{author}{\bibinfo{person}{Naiqian Zheng}, \bibinfo{person}{Mengqi
  Liu}, \bibinfo{person}{Ennan Zhai}, \bibinfo{person}{Hongqiang~Harry Liu},
  \bibinfo{person}{Yifan Li}, \bibinfo{person}{Kaicheng Yang},
  \bibinfo{person}{Xuanzhe Liu}, {and} \bibinfo{person}{Xin Jin}.}
  \bibinfo{year}{2022}\natexlab{}.
\newblock \showarticletitle{Meissa: Scalable Network Testing for Programmable
  Data Planes}. In \bibinfo{booktitle}{\emph{Proceedings of {ACM} {SIGCOMM}
  2022 Conference, {SIGCOMM} 2022}}. \bibinfo{publisher}{{ACM}},
  \bibinfo{pages}{350--364}.
\newblock


\end{thebibliography}

\end{document}